\def\ba{\bar a}
\def\ha{{1\over 2}}
\def\tt{\tilde t}

\def\be{\begin{equation}}
\def\ee{\end{equation}}
\def\te{\end{equation}}
\def\bea{\begin{eqnarray}}
\def\ba{\begin{eqnarray}}
\def\eea{\end{eqnarray}}
\def\ea{\end{eqnarray}}
\def\tea{\end{eqnarray}}
\def\nn{\nonumber}

\newcommand{\eqn}[1]{(\ref{#1})}

\def\(#1){(\ref{#1})}

\newskip\humongous \humongous=0pt plus 1000pt minus 1000pt

\newif\ifdtup

\documentclass[a4paper]{jpconf}
\usepackage{graphicx}

\makeatletter                    
\@addtoreset{equation}{section}  
\makeatother                     

\begin{document}

\title{Macroscopic quantum phenomena \\ from the large N perspective}

\author{C H Chou$^1$, B L Hu$^2$ and Y Suba\c{s}\i$^2$}
\address{$^1$Department of Physics, National Cheng Kung                                 
University, Tainan, Taiwan 701 and \\ National Center for                        
Theoretical Sciences (South), Tainan, Taiwan 701 }
\address{$^{2}$ Joint Quantum Institute and Maryland Center for Fundamental Physics,\\ University of Maryland, College Park, Maryland 20742, USA}


\ead{blhu@umd.edu}

\begin{abstract} Macroscopic quantum phenomena (MQP) is a relatively new research venue, with exciting ongoing experiments  and bright prospects, yet with surprisingly little theoretical activity. What makes MQP intellectually stimulating is because it is counterpoised against the traditional view that macroscopic means classical. This simplistic and hitherto rarely challenged view need be scrutinized anew, perhaps with much of the conventional wisdoms repealed. In this series of papers we report on a systematic investigation into some key foundational issues of MQP, with the hope of constructing a viable theoretical framework for this new endeavour. The three major themes discussed in these three essays are the large N expansion, the correlation hierarchy and quantum entanglement for systems of `large' sizes, with many components or degrees of freedom.  In this paper we use different theories in a variety of contexts to examine the conditions or criteria whereby a macroscopic quantum system may take on classical attributes, and, more interestingly, that it keeps some of its quantum features. The theories we consider here are, the $O(N)$ quantum mechanical model, semiclassical stochastic gravity and gauge / string theories; the contexts include that of a `quantum roll' in inflationary cosmology,  entropy generation in quantum Vlasov equation for plasmas, the leading order and next-to-leading order large N behaviour, and hydrodynamic / thermodynamic limits. The criteria for classicality  in our consideration include the use of uncertainty relations, the correlation between classical canonical variables, randomization of quantum phase, environment-induced decoherence, decoherent history of hydrodynamic variables,  etc. All this exercise is to ask only one simple question: Is it really so surprising that quantum features can appear in macroscopic objects? By examining different representative systems where detailed theoretical analysis has been carried out, we find that there is no a priori good reason why quantum phenomena in macroscopic objects cannot exist.

\end{abstract}

\centerline{\small {\it-- This is a later version than the Feb. 24, 2011 version appearing in J. Physics (Conf. Series) for DICE 2010 meeting.}}

\section{Quantum / classical, micro / macro}
\label{sec:Q/C,M/M}

There are many ways to deal with the issue of quantum-classical correspondence \cite{QCCDrexel}. In the most common and traditional view the classical limit corresponds to  $\hbar \rightarrow 0$, or, invoking the Bohr correspondence principle, the principal quantum number of a system $n \rightarrow \infty $, or regarding the coherent state as the `most classical' quantum state, or the Wigner function as the `closest to classical' distribution. Less precise criteria also abound, such as the loose concept that a system at high temperature behaves classically, or viewing the thermodynamic / hydrodynamic limits (of a quantum system) as classical. (For a description of the various criteria, see, e.g., \cite{HuZhaUnc}). There are holes in almost all of the above common beliefs.  A more sophisticated viewpoint invokes decoherence, the process whereby a quantum system loses its coherence (measured by its quantum phase information) through interaction with its environment \cite{envdec}.   In this work we examine an alternative perspective, as the folklore goes,  that quantum pertains to the small (mass, scale) while classical to the large (size, multiplicity).  This common belief now requires a much closer scrutiny in the face of new challenges from macroscopic quantum phenomena (MQP), viz, quantum features may show up even at macroscopic scales. A common example is superconductivity where the Cooper pairs can extend to very large scales compared to interatomic distances and Bose-Einstein condensate (BEC) where a large number N of atoms occupy the same quantum state, the N-body ground state. Other examples include nanoelectromechanical devices \cite{nem} where the center of mass of a macroscopic classical object, the cantilever, obeys a quantum mechanical equation of motion. Experiments to demonstrate the quantum features such as the existence of interference between two macroscopic objects have been carried out, e.g., for $C^{60}$ molecules passing through two slits \cite{Arndt} or proposed, e.g., for two mirrors \cite{Marshall,EntSQL}.

A most direct account of the  difference between the microscopic and the macroscopic behaviours of a quantum system is by examining  N, the number of physically relevant (e.g. for atomic systems, forgetting about the tighter-bound substructures) quantum particles or components in a macroscopic object.  One may ask:  At what number of N will it be suitable to describe the object as mesoscopic with  qualitatively distinct features from microscopic and macroscopic? In the  case of classical equilibrium statistical mechanics this issue is easily addressed by use of the (grand) canonical ensemble, and such inquiry is answered by understanding how the (grand) partition function behaves as a function of N and temperature $T$ (and chemical potential $\mu$) all the way to the thermodynamic limit. In classical nonequilibrium statistical  mechanics, this issue underlies the derivation from molecular Hamiltonian dynamics  the thermodynamic and kinetic properties (such as transport functions) of a gas of N molecules, and their dynamics, which possess salient dissipative and time-asymmetry features nonexistent in the microscopic dynamics. Boltzmann answered this question with great valour and magnificent success, bringing forth also the issue of time-arrow and providing  a molecular dynamics basis for the Second Law. To perform a quantitative analysis of such issues one needs to work with  kinetic and stochastic equations for the N particles, the Boltzmann equation -BBGKY hierarchy for (effectively) closed systems in the case where there is no distinguished party, and when there is, the Langevin equation for open systems.  In the recent decade significant advances have been made in providing a  molecular dynamics basis to the foundations of thermodynamics  \cite{Dorfman}, relating the macroscopic thermodynamic behaviour of a gas to the chaotic dynamics of its molecular constituents.  One could even calculate the range in the number of molecules where a microscopic system begins to acquire macroscopic behaviour and hence identify the approximate boundaries of mesoscopia \cite{Gaspard}.

For quantum systems one needs to deal with additional concerns of quantum coherence and entanglement which are critically important issues in quantum information processing (QIP) \cite{QIPsi}. A fundamental issue in QIP is how the performance of a quantum information processor alters as one scales the system up. This dependence on N  is known as the ``scaling" problem \cite{DVcriteria}. We will defer considerations of quantum entanglement of macroscopic objects to a third paper in this series after we have a chance to explore how quantum correlations and fluctuations impact on MQP in the second paper using the n particle-irreducible (nPI) representation. There are many important and interesting issues of MQP,  one subset of special interest to us is how quantum expresses itself in the macroscopic domain since usually macro conjures classicality. Thus even the simplest yet far from naive question need be reconsidered properly. For example, why is it that an ostensibly macroscopic object such as a cantilever should follow a quantum equation of motion.  This center-of-mass axiom is implicitly assumed in many descriptions of MQP but rarely justified. The conditions upon which this can be justified are provided by two of us and a different third author \cite{CHY} with the derivation of a master equation for N harmonic oscillators (NHO) in a finite temperature harmonic oscillator bath.


In this paper we ask the question: When will a quantum system of N particles or components  begin to acquire or show classical features. We shall present an array of physical examples to illustrate how this comes about, and glean from them 1) how classicality is defined, 2) How the meaning of classicality changes in different types of theories?  For 1) the criteria mentioned at the beginning and expounded further in the next section will be invoked in several concrete examples, such as, how the inflaton dynamics determines when the inflationary universe begins to behave classically, from the simplest criteria (invoking the uncertainty principle) to the more involved (correlation between canonical variables) and sophisticated (decoherence). Other criteria studied are randomization of quantum phase, entropy generation, decoherence of hydrodynamic variables due to the existence of conservation laws.  For 2) we give a detailed study of a) the O(N) model: an N component quantum scalar field $\phi^4$ with $\lambda \phi^4$ self interaction (in section~\ref{sec:ONlargeN}) (see, e.g., \cite{CHKMPA,CHKM97,HuRam97}), the quantum mechanical version of which describes a system of N interacting quantum harmonic oscillators, as it is the common underlying structure of many physical examples, where some claims of macrocity (being macroscopic, for lack of a better noun) and classicality originate.  For our purpose here it is sufficient to work with a zero-dimensional field theory, i.e., a quantum mechanical model --  not only can one shred off the shrouded field theory technicalities but it is more directly related to atomic-optical and condensed matter set-ups for easier experimental comparisons. b)  semiclassical stochastic gravity \cite{stogra}, and c) gauge /  string theory (in section~\ref{sec:NLOLN}). We use some well-known results in quantum field theory - the large N expansion -- to consider mean field behaviour and under what conditions the semiclassical limit of a quantum theory will be reached  at the leading order (LO) large N expansion, and the role of quantum fluctuations at the next-to-leading-order (NLO) large N expansion.

Both the O(N) theory  and semiclassical gravity show semiclassical behaviour when N becomes very large. For the former the leading order large N expansion yields a mean field theory with Vlasov equation which is time-reversal invariant. We use this model as an example to expound these issues in section~\ref{sec:LOLN}. The next-to-leading order (NLO) expansion begins to show dissipative behaviour in the two particle irreducible (2PI) representation \cite{CJT}. Using the closed-time-path (CTP) 2PI NLO effective action an H-theorem can be shown to exist for the quantum mechanical O(N) model \cite{CH03}.  In the latter example, perturbative quantum gravity interacting with an N component quantum scalar field emerges as the semiclassical limit when N becomes very large \cite{HarHor}. The equations of motion (semiclassical Einstein equation) is time-reversal invariant. The NLO expansion yields stochastic gravity \cite{RouVerLN} which has both dissipative and fluctuation features.  These prior results provide a stimulus for us to inquire about the quantum-classical correspondence issue in an alternative way, i.e., instead of using the loop expansion (in orders of $1/ \hbar$) where quantum features reveal at successively higher loop orders as corrections to the classical solution, we   use the large N expansion, where N provides a quantitative measure of the `magnitude' or `complexity' of a quantum system, to examine how the semiclassical limit of a quantum theory is reached. We shall rely heavily on the results obtained in \cite{MACDH,MDC,CHthermalize}  to illustrate these crucial aspects in MQP. However, as the examples in gauge and string theory show \cite{Hooft,Witten79,Maldacena} (section~\ref{sec:NLOLN})  the leading order large N does not necessarily correspond to a classical limit. It may correspond to a thermodynamics limit instead. These are the purposes for our invoking different theories to illustrate different criteria and see their fine and not so fine distinctions.

Perhaps it is necessary to make this remark here before we describe the contents of this paper: many ingredients in our discussions are well-known to different communities, such as the large N expansion in field theory and critical phenomena,  dissipation and fluctuations in nonequilibrium statistical mechanics, stochastic equations in quantum open systems, hydrodynamic variables and quasi-classical domains in decoherent histories, etc. We see this paper's contribution is a modest one: discerning the physical essence of apparently similar technology and sharpening the often mixed-up physical meanings while focusing on the issues interspersed between the macro and the quantum. We hope to use these well-known parts to elucidate some lesser known, even misunderstood facts, and to shape a new perspective, as a first step towards constructing a theoretical foundation for MQP.  \footnote{Because we want to be as concrete as possible in making our points we decided to extract technical details verbatim from some representative papers on the specific topics, instead of grossing over and talking abstractly. For this we show our appreciation to the authors of these papers for their lucid presentations which facilitate our illustrative purpose.
An apology to the readers goes along with our appreciation to the authors: To experts in field theory, string theory, gravity and cosmology, the themes in this paper may be too familiar to them and the inclusion of details may thus make the contents overladen. This is because we need to consider readers from the atomic, optical and condensed matter community where MQP experiments and theories are currently undertaken.}

Readers familiar with these theories should just skip over the technical details in those parts and concentrate on how they bring out the physical issues raised.   In section~\ref{sec:LNQC} we display an array of criteria behind the common beliefs pertaining to quantum, classical and macro, draw the linkages between them, differentiate seemingly equivalent criteria, and issue some warnings to over-generalizations.  In section~\ref{sec:ONlargeN} we  present a) the $O(N)$  model,  give a quick summary of the large N expansion method, leading to a set of coupled equations where one can see the variation with N in physical quantities of interest.   In section~\ref{sec:LOLN} we use three examples to describe how a macroscopic quantum system acquires classical features: 1) The so-called `quantum roll' model of inflationary cosmology \cite{GuthPi} based on the $O(N)$ model, how stringent the different classicality criteria are: amongst them the use of the uncertainty relation, correlation in conjugate variables, and environment-induced decoherence. 2) Quantum Vlasov equation (QVE)  as an example of a mean field theory obtained from the leading order large N (LOLN) expansion of the $O(N)$ model. Following \cite{KME} we express QVE  as a coupled set of equations, one for the number of particles, the other for their correlations which contain quantum phase relations. This enables one to see how coarse-graining over the phase information brings about entropy generation which is related to uncertainty which can be used to quantify classicality. We also mention how in the decoherent history conceptual framework quasi-classical domains can appear owing to the existence of conservation laws for hydrodynamic variables, thus highlighting the role of emergence and importance of collective behaviour in MQP. In section~\ref{sec:NLOLN} we consider the next-to-leading order large N expansion and try to identify the role of quantum fluctuations in MQP, using examples given above to make explicit the conditions a macroscopic system can shred off or retain its quantum features. We use two more illustrative theories here, b) semiclassical  stochastic gravity and c) gauge / string theories to examine their LOLN and NLO dynamics and explore their classical versus quantum behaviours.  We learned that there is no a priori reason why a quantum system with large N need to be classical.  We discuss the differences in physical meaning between a mean field and a classical field, and under what conditions the former leads to the latter. Only for the class of Gaussian theories like the O(N) model where the mean field theory obtained from the LOLN expansion is equivalent to  the classical theory would it make sense to think of deviations (from LO expansion) from the mean  as providing the quantum corrections which then allows  one to use large N to address the quantum- classical correspondence issue.   In section~\ref{sec:conclusion} we summarize what we have learned in this intellectual exercise and make some general observations. The overall lesson we learned  is that we need to be very careful in finding the conditions and formulating criteria for a system with large N to behave classically.  To be safe, one should think of all systems as quantum intrinsically at all orders of N, and not to identify mean field as classical without knowing the structure and behavior of the theories we are talking about.

\section{Large N in relation to quantum and classical}
\label{sec:LNQC}

In this program we have set up for ourselves we need to clearly distinguish the relation between micro-macro / quantum-classical as well as the thermo-hydro limits. The micro-macro correspondence underlies the basic theme of statistical mechanics which has been investigated for centuries, and although  mesoscopic physics has been with us for three decades many key issues bearing on quantum correlations and fluctuations remain a challenge today.   The quantum-classical transition /correspondence has been studied in earnest in the last two decades and we have gained much understanding of the role environments play in bringing about a quantum system's classical behaviour, but the issue of quantum and macro remains little explored, since most common beliefs associate micro with quantum and macro with classical. MQP is a strong statement that this need not be the case.  We ought to scrutinize all conventional assumptions and reexamine every single concept and criterion before reconnecting them into a new conceptual framework to meet this new challenge.

As a background for motivation and a collection of key points we list below the common beliefs for a macroscopic object to assume classical behaviour. We pay special attention to the relation of large N, quantum and classicality.

\subsection{$\hbar$, loop expansion and classicality}

\begin{itemize}

\item 
Why is it that $\hbar \rightarrow 0$ is often viewed as classical?
This can be easily seen from the path integral formalism: $1/\hbar$
multiplies the action in the path integral, which has no further
dependence on $\hbar$. The dominant contribution to the path
integral comes from the path which extremizes the action,
i.e., the classical trajectory. The steepest descent
method becomes exact when $\hbar \rightarrow 0$. It is in this sense that taking $\hbar \rightarrow 0$ give rise to classical physics.

Suppose the macroscopic object is made up of a space-time inhomogeneous
boson condensate. Its  classical behaviour manifests itself via the vacuum expectation value
of a scalar field $\psi$:
\begin{eqnarray*}
\phi(\vec{x},t) = \langle 0| \psi(\vec{x},t) |0 \rangle
\end{eqnarray*}
where $|0 \rangle$ is the vacuum with the space-time inhomogeneous boson condensate creating the macroscopic object. Note that
$\psi^n$ contains not only $\phi^n$, but also the product of
normal product terms. Rearranging a product of normal product
terms into a normal product creates c-numbers due to the
contraction of creation and annihilation operators, while each
contraction creates a c-number of order $\hbar$, namely,
\begin{eqnarray*}
\langle 0| \psi(\vec{x},t)^n |0 \rangle = \phi(\vec{x},t)^n
+O(\hbar).
\end{eqnarray*}
Thus only when the contractions which produce the loops in the Feynman
diagrams are ignored can one replace $\langle 0| \psi^n |0
\rangle$ by $\phi^n$. This gives the tree level approximation which is usually termed ``classical".
\end{itemize}

\subsection{Large N, mean field and fluctuations}

\begin{itemize}

\item \textit{Fluctuations to mean}:  Bohr's correspondence principle  states that the classical behaviour of  a quantum system with $n$ quantum levels is established when the quantum fluctuation $ \Delta n$ is much smaller than the average quantum number $n$, i.e. $\Delta n/n \ll 1$. This condition does not depend on $\hbar$, but is controlled by the number of participating particles or quanta.

\item \textit{Large $N$ as a mean field}:
The limiting theory when $N \rightarrow \infty$ in a leading order large N expansion is usually referred to as a \textit{mean field theory} in that contributions from fluctuations are maximally suppressed. To see this note that in every theory known to have a sensible large $N$ limit,
the vacuum expectation of any product of operators, $\hat{A}\hat{B}$, satisfies the factorization relation
\begin{eqnarray*}
\langle \hat{A} \hat{B} \rangle = \langle \hat{A}  \rangle \langle
\hat{B} \rangle +O(1/N).
\end{eqnarray*}
(This means that the disconnected Feynman graphs always dominate).
Therefore the variance of any operator vanishes as $N\rightarrow
\infty$,
\begin{eqnarray*}
\lim_{N\rightarrow \infty}(\langle \hat{A}^2  \rangle - \langle
\hat{A}  \rangle^2 ) = 0.
\end{eqnarray*}
This means that the fluctuations become negligible when $N \rightarrow \infty$.

\item {\it $\hbar \rightarrow 0$ and $N \rightarrow \infty$ are different limits.}
The loop expansion and large N expansion are very different.
In QCD, $N$ determines the number of degrees of
freedom ($N^2 -1$ gluons), therefore the limit $N \rightarrow
\infty$ resembles the infinite volume limit of a two-dimensional
lattice system. This is different for theories involving
$N$-component vector fields such as the $O(N)$ model described in section~\ref{sec:ONlargeN}. There, the $\hbar \rightarrow 0$ limit gives the complete classical theory whereas $N \rightarrow \infty$ is the gaussian approximation of a quantum theory, for which classical and quantum dynamics are identical. So even for this class of theories these two limits are, strictly speaking, different.
\end{itemize}

\subsection{Large N, classical or thermodynamic limits?}

\begin{itemize}

\item \textit{Is the mean field a classical field?}
There is a common misunderstanding which equates mean field with classical field, whence  $N \rightarrow \infty$ limit is viewed as a condition of classicality. But as pointed out by Habib \cite{HabibPRL,Habib08}(and reference therein) it is only for Gaussian systems such as  the O(N) QM model that this is true. It is not true, e.g., for gauge fields, as shown below.


\item \textit{Large $N$ as  thermodynamic limit?}   Coleman in his 1979 Erice Lectures \cite{Coleman}, perhaps the most commonly read introduction to large N field theory, drew a parallel with the classical limit: ``There is a classical gauge field configuration, which I will call the master field, such that the large-N limits of all gauge invariant Green functions are given by their values at the master field." Along this line Witten in his 1979 Cargese lectures  \cite{Witten79} presented an argument for the existence of a master field in the large-$N$ limit of QCD.  However,  O. Haan \cite{Haan} pointed out that a crucial assumption Witten made, that the expectation value of any $U(N)$ symmetric operator of the form $f(A) = {\rm Tr} [A(x_1)...A(x_l)] $ where $A$ are $N \times N$ matrix fields, factorizes, i.e., $<f_1(A) ... f_k(A)> = \Pi^k_{i=1} <f_i(A)> [1+ O(1/N^2)]$, does not hold. He showed that for the two matrix models this `masterfield' does not exist. Instead of the classical limit Haan argued that the factorization of expectation values of invariant operators is a property analogous to the vanishing of fluctuations for macroscopic observables in the thermodynamic limit. We will return to this issue in section 5.3.

\item \textit{Hydrodynamics and thermodynamics limits.} See section~\ref{subsec:DecHyd} on this issue.

\end{itemize}

\subsection{Coherent state and classicality}

\begin{itemize}

\item Yaffe in \cite{Yaffe} stated that to view Large $N$ limits as classical
mechanics  requires 4 assumptions. If
these assumptions are satisfied, one can generate a natural set of
generalized coherent states. These coherent states may then be
used to construct a classical phase space, derive a classical
Hamiltonian, and the resulting classical dynamics is equivalent to
the limiting form of the original quantum dynamics.

We note that by way of a microscopic model in the open quantum system framework Zurek, Habib and Paz  \cite{ZHP} studied environment-induced decoherence and showed that there is a tendency for a quantum system to evolve to a coherent state which is commonly viewed as the `most classical' quantum state. This is related to the uncertainty relation criterion for classicality as illustrated in \cite{HuZhaUnc,AnaHal}.

\end{itemize}

\section{Quantum mechanical O(N) model: large N expansion}
\label{sec:ONlargeN}

Large N expansion of the O(N) model in 3+1 dimensional quantum field theory  has been investigated extensively in the literature. The results of immediate relevance for our purpose are from the work of  Cooper, Dawson, Habib, Kluger, Mihaila, Mottola, Paz et al. For the O(N) quantum mechanical model describing the dynamics of a system of N nonlinear oscillators we follow the treatment of \cite{MACDH,MDC},  while relegating the 2PI aspects \cite{CH03,CHthermalize} to our sequel paper. For this model  numerical integration valid for all N is possible with the quantum roll initial condition \cite{GuthPi} which provides a measure of how accurate the large N (LO and NLO) expansion is as a function of N. We will make use of these results to analyze the quantum-classical problem for large N quantum systems.

\subsection{O(N) model}

The Lagrangian for the $O(N)$ model in quantum mechanics is given by:
\begin{equation}
   L(x,\dot{x})
   =
   \frac{1}{2} \sum_{i=1}^N \dot{x}_i^2 - V(r)  \>,
   \label{eq:Lagmod}
\end{equation}
where $r^2 = \sum_{i=1}^N x_i^2 $ and $V(x)$ is a potential of the form
\begin{equation}
   V(r)  =  \frac{g}{8 N} \,
      \left  (
         r^2  -  r_0^2
      \right )^2
   \label{eq:classi}
\end{equation}
where $g$ is the self-coupling constant (related to the $\lambda$ in its field theory origin) and $r_0$ is where the potential goes to zero, $V(r_0)=0$.
The time-dependent Schr\"odinger equation for this problem is given
by:
\begin{equation}
\rmi \, \frac{\partial \psi(x,t)}{\partial t} \>=
\left  \{ - \frac{1}{2} \sum_{i=1}^N \frac{\partial^2}{\partial x_i^2}
+ V (r) \right \} \, \psi(x,t).
\end{equation}
The initial conditions for the quantum roll problem
allow a numerical solution for all $N$, as only the radial
part of the wave function enters. With the exact solution available one can examine the validity of the results from a large-$N$ expansion.

Assuming a solution of the form
\begin{equation}
   \psi(r,t) = r^{(1-N)/2} \phi(r,t),
\end{equation}
the time dependent Schr\"odinger equation for $\phi(r,t)$ reduces
to \cite{BlazotRipka}:
\begin{equation}
   \rmi \, {\partial \phi(r,t) \over \partial t} \>=\left \{ -
      \frac{1}{2}\frac{\partial^2}{\partial r^2}
      + U(r) \right \} \, \phi(r,t)
   \label{eq:redham}
\end{equation}
with an effective one dimensional potential $U(r)$ given by
\begin{equation}
   U(r)
   =
   \frac{(N-1)(N-3)}{8 \, r^2}
   +
   \frac{g}{8N} \, \left ( r^2 - r_0^2 \right )^2
   \>.
\label{eq:Uofr}
\end{equation}
Upon the rescaling
\begin{equation}
   r^2 = N y^2 \>, \qquad r_0^2 = N y_0^2 \>.
   \label{eq:yscaling}
\end{equation}
it becomes
\begin{equation}
   u(y,N)
   =
   \frac{U(y)}{N}
   =
   \frac{(N-1)(N-3)}{8 N^2 \, y^2} +
   \frac{g}{8} \, ( y^2 - y_0^2 )^2  \>,
   \label{eq:Uscaled}
\end{equation}
corresponding to the new Schr\"odinger equation,
\begin{equation}
   \rmi \, {\partial \phi(y,\tilde{t}) \over \partial \tilde{t}} \>=
\left \{
      - \frac{1}{2N^2}
      \frac{\partial^2}{\partial y^2}
      + u(y,N)
   \right \} \, \phi(y,\tilde{t})
   \label{eq:schroii}
\end{equation}
where $\tilde{t} = N \, t$.

To implement the large N expansion, following the original work of \cite{CJP}, we introduce a composite field $\chi$  and add a constraint term
\begin{equation}
   \frac{N}{2 g} \,
      \left  [
         \chi -
         \frac{g}{2 N} ( r^2 - r_0^2 )
      \right ]^2  \>,
\end{equation}
to  the original Lagrangian which yields an equivalent Lagrangian,
\begin{equation}
   L'(x,\dot{x},\chi) =
      \sum_i
         \frac{1}{2} \left  (
                        \dot{x}_i^2 - \chi x_i^2
                     \right )
       + \frac{r_0^2 }{2} \chi
       + \frac{N}{2 g} \chi^2 \>.
\label{eq:LLN}
\end{equation}
One can also glean off the results from the corresponding field theory treatment in \cite{CHKMPA} (by specializing to $0+1$ dimensions, and
replacing $\phi_a(t) \rightarrow x_i(t)$, $\mu^2 \rightarrow -\frac{r_0^2 g}{2N}$, and
$\lambda \rightarrow g$.)

\subsection{Large N expansion}

The generating function $Z[j,J]$ is given by the path integral over
the background fields $x_i(t)$:
\begin{eqnarray*}
   Z[j,J]
   = \rme^{\rmi W[j,J]}
   &=&
   \int {\rm d}\chi \ \prod_i {\rm d} x_i \,
        \exp
        \Bigl \{
           \rmi \, S[x,\chi;j,J]
        \Bigr \}  \>,
   \\
   S[x,\chi;j,J]
   &=&
    \int_{\cal C} {\rm d}t \,
    \Bigl \{
       L' + \sum_i j_i x_i + J \chi
    \Bigr \}  \>.
\end{eqnarray*}
The effective action, to order $1/N$, is obtained by integrating the
path integral for the generating functional for the Lagrangian
(\ref{eq:LLN}), over the $x_i$ variables, and approximating the
integral over $\chi$ by the method of steepest descent (keeping terms
up to order $1/N$).  A Legendre transform of the resulting generating
functional then yields the effective action, which we find to be:
\begin{eqnarray}
\Gamma[q,\chi] &=& \int_{\cal C} \rmd t \,
       \biggl \{
          \frac{1}{2} \sum_i \,
          \Bigl [
             \dot{q}_i^2(t) - \chi(t) \,  q_i^2(t)
          \Bigr ]
          +
          \frac{\rmi}{2} \sum_i \,
          \ln \, [ G_{ii}^{-1}(t,t) ]
   \nonumber \\
   && \qquad
          + \frac{r_0^2}{2} \, \chi(t)
          + \frac{N}{2 g} \, \chi^2(t)
          + \frac{\rmi}{2} \ln [ D^{-1}(t,t) ]
       \biggr\}
   \>,
\label{eq:effaction}
\end{eqnarray}
where the integral is over the close time path $\cal C$, discussed in \cite{CHKMPA} and $q(t) = \langle x_i(t) \rangle$.  Here
$G^{-1}_{ij}(t,t')$ and $D^{-1}(t,t')$ are the lowest order in $1/N$
inverse propagators for $x_i$ and $\chi$, given by
\begin{eqnarray*}
   G^{-1}_{ij}(t,t')
   & = &
      \left  \{
         \frac{ {\rm d}^2 }{ {\rm d} t^2 } + \chi(t)
      \right \} \, \delta_{\cal C} (t,t') \, \delta_{ij}
   \equiv
   G^{-1}(t,t') \, \delta_{ij}
   \>, \\
   D^{-1}(t,t')
   & = &
      - \frac{N}{g} \delta_{\cal C}(t,t')
      -  \Pi(t,t')
   \>,
\end{eqnarray*}
where
\begin{eqnarray}
   \Pi(t,t')
   & = &
      - \frac{\rmi}{2} \, \sum_{i,j} G_{ij}(t,t') \, G_{ji}(t',t)
   \nonumber \\ &&
      + \sum_{i,j} \, q_i(t) \, G_{ij}(t,t') \, q_j(t') \>.
   \label{eq:PilrgNdef}
\end{eqnarray}
Here $\delta_{\cal C} (t,t')$ is the closed time path delta function.

The equations of motion for the background fields $q_i(t)$, to order
$1/N$, are
\begin{equation}
   \left  \{
      \frac{{\rm d}^2 }
           {{\rm d} t^2 }
      + \chi(t)
   \right\} q_i(t)
   + \rmi \, \sum_{j} \int_{\cal C} {\rm d}t' \,
         G_{ij}(t,t') \, D(t,t') \, q_j(t')
   = 0
   \>,
\label{eq:phieom}
\end{equation}
with the gap equation for $\chi(t)$ given by
\begin{equation}
   \chi(t)
   =
   - \frac{g}{2N} \, r_0^2 +
     \frac{g}{2 N} \sum_i
      \left [
         q_i^2(t) \, + \, \frac{1}{\rmi} \, {\cal G}_{ii}^{(2)}(t,t)
      \right ]
   \>.
\label{eq:Chieqn}
\end{equation}

The next-to-leading order $x_i$ propagator ${\cal G}_{ij}^{(2)}(t,t')$
and self energy $\Sigma_{ij}(t,t')$ to order $1/N$ are
\begin{eqnarray}
   &&
   {\cal G}_{ij}^{(2)}(t,t')
   \ = \
   G_{ij}(t,t') \label{eq:Gfull}
   - \, \sum_{k,l}
      \int_{\cal C} {\rm d}t_1 \, \int_{\cal C} {\rm d}t_2 \,
      G_{ik}(t,t_1) \, \Sigma_{kl}(t_1,t_2) \, G_{lj}(t_2,t') \>,
   \nonumber \\
   &&
   \Sigma_{kl}(t,t')
   \ = \
   \rmi \, G_{kl}(t,t') \, D(t,t')
     - q_k(t) \, D(t,t') \, q_l(t')
   \>.
\end{eqnarray}
(compare these equations with (2.18--2.22) of \cite{CHKMPA}.)

The actual equation for $\cal G$ which follows
from the effective action differs from (\ref{eq:Gfull}) in that the
final G in the integral equation is replaced by the full $\cal G$. This
leads to a partial resummation of the 1/N corrections which guarantees positivity of $\langle x^2(t)\rangle$ (but not the full positivity for
the density matrix).

In order to solve for $D(t,t')$, we first write
\begin{equation}
   \frac{N}{g} \, D(t,t') \ = \
      - \, \delta_{\cal C}(t,t')
      \, + \,
      \frac{N}{g} \, \Delta D(t,t')
   \>,
\label{eq:Dsubst}
\end{equation}
then $\Delta D(t,t')$ satisfies the integral equation,
\begin{equation}
   \frac{N}{g} \, \Delta D(t,t')
   \ = \
   \frac{g}{N} \, \Pi(t,t')
   - \int_{\cal C} {\rm d}t'' \,
            \Pi(t,t'') \, \Delta D(t'',t')
   \>,
\label{eq:Dtileqn}
\end{equation}
(Compare with (2.13--2.16) of Ref.~\cite{CHKMPA}.)

\subsection{Effective potential}

As pointed out by Mihaila et al \cite{MACDH} the effective potential in the large $N$ approximation has been
previously obtained by Root \cite{Root} to order $1/N$. We continue to follow their treatment. From the action \ref{eq:effaction}, they found that \footnote{When $x_i$ and $\chi$ are independent of time, one can ignore the closed time path ordering and
use Fourier transforms, passing the poles by using the Feynman contour.}
\begin{eqnarray}
   \lefteqn{
   V_{\rm{eff}}^{[1]}(r,\chi)  =
   \frac{N \chi}{g} \,
   \left ( \mu^2 - \frac{\chi}{2} \right )
   +
   \frac{1}{2} \, \chi \, r^2 }
   \label{eq:Veff} \\
   &&
   +
   \frac{N}{2} \int \frac{\rmd k}{2\pi \rmi} \, \ln [ \tilde G^{-1}(k) ]
   +
   \frac{1}{2} \int \frac{\rmd k}{2\pi \rmi} \, \ln [ \tilde D^{-1}(k) ]
   \>,
   \nonumber
\end{eqnarray}
where $\chi$ satisfies the requirement
\begin{eqnarray}
   \frac{\partial}{\partial \chi} V_{\rm{eff}}(r,\chi)
   =
   0 \>.
   \label{eq:pVeffpChi}
\end{eqnarray}
Using the rescaled $y$ variables defined in (\ref{eq:yscaling}) the effective potential (\ref{eq:Veff}) becomes
\begin{eqnarray}
   {V_{\rm{eff}}^{[1]}(y,\chi)  \over N}
   & = &
   \frac{\chi}{2} \, \bigl ( y^2 - y_0^2 \bigr ) -
   \frac{\chi^2}{2 \, g} +
   \frac{\sqrt{\chi}}{2}
   \label{eq:Veff_sec} \\
   && \qquad {}+
   \frac{1}{2 \, N} \,
      \bigl ( m_{+} + m_{-} - 3 \, \sqrt{\chi} \, \bigr )  \>.
   \nonumber
\end{eqnarray}
where $m_{\pm}^2=b\pm \sqrt{b^2-c}$, with
\begin{eqnarray}
b
&=&
\frac{5}{2}\chi+
\frac{g}{2}\left( y^2+\frac{1}{2\sqrt{\chi}}\right)\\
c
&=&
4\chi^2+
g\left( 4y^2\chi +\frac{1}{2}\sqrt{\chi}\right)
\label{eq:m+-}
\end{eqnarray}
The gap equation which determines $\chi$ follows from
(\ref{eq:pVeffpChi})
\begin{eqnarray}
   \chi
   & = &
   \frac{g}{2} \, \bigl ( y^2 - y_0^2 \bigr ) +
   \frac{g(N - 3)}{4 \, N \,\sqrt{\chi} } +
   \frac{g}{2\,N} \,
      \frac{\partial (m_{+} + m_{-})}{\partial \chi} \>.
   \label{eq:chi_sec}
\end{eqnarray}

To leading order in the large $N$ expansion, (\ref{eq:Veff_sec}),
(\ref{eq:chi_sec}) reduce to the parametric set \footnote{These equations (\ref{eq:Veff_sec}) and (\ref{eq:chi_sec}) agree with Root  \cite{Root},
however he used the leading order expression for $\chi$ in
(\ref{eq:Vchi0}), rather than the full~$\chi$ of (\ref{eq:chi_sec}).}
\begin{eqnarray}
   {V_{\rm{eff}}^{[0]}(\chi) \over N}
   & = &
   \frac{\chi^2}{2g} + \frac{ \sqrt{\chi}}{4} \>,
   \nonumber \\
   y^2(\chi)
   & = &
   y_0^2 + \frac{2}{g} \, \chi - \frac{1}{2 \, \sqrt{\chi}}
   \>. \label{eq:Vchi0}
\end{eqnarray}

These snugly placed equations are very useful for our purpose of analyzing the aforementioned issues in MQP. A result of significance is finding the lower bound on N whereby the NLO results give a good approximation to the exact. How does this bear on the issues at hand, namely, quantum behavior and classicality in macroscopic systems? We will discuss this in section 5.1.

In the next section we will focus on the leading order results which are much simpler but still provides useful insight on the macro-quantum relation.



\section{Number - phase / correlation, coarse-graining / uncertainty - entropy, decoherence / classicality and hydrodynamics}
\label{sec:LOLN}

In this section while staying at the leading order large N (LOLN) level we adopt another angle to analyze mean field theory and classicality. Using the uncertainty relation to find out when a quantum system begins to show classical behaviour is shown to be inadequate, as we will illustrate with an example of quantum roll in the inflationary universe. Decoherence and correlation are needed. In the second example we use a theory shown to be obtainable as the LOLN of an O(N) model which is time-reversal invariant and examine another set of criteria for classicality, that of quantum phase information and entropy generation.
Although a system with no quantum phase information is usually identified as classical, yet in essence this is not the case. Fluctuations in the number density carry phase information which signifies its quantum origin. One needs to ensure that phase information at all orders vanish for all times to be able to say so, but this is not natural and not easily implementable.
In the third subsection we discuss  classicality from the decoherent history viewpoint, that hydrodynamic variables are more readily decohered because of the existence of conservation laws for these variables. We also mention why thermodynamic and classical limits are not identical.

\subsection{Criteria of classicality: uncertainty relation, decoherence and correlation in a quantum roll}

One commonly used signifier of when a quantum system begins to show classical features is the use of the uncertainty relation to distinguish these two regimes. (See \cite{HuZhaUnc}.) A concrete example with some importance in early universe cosmology is Guth and Pi (GP)'s  \cite{GuthPi} use of a quantum mechanical model for the description of the inflaton dynamics of a scalar field (the inflaton). They solve for the evolution of the wave function in a de Sitter universe to illustrate how the inflationary transition takes place. The O(N) model has been used in Friedmann universe models by Boyanovsky \cite{Boy}, Cooper et al \cite{CooperPi86} and many others.  The scalar field starts at the top of  an inverted harmonic oscillator potential and `rolls down'; this so-called `quantum roll'  initial condition is in fact used in the results reported in the previous section by Mihaila et al. We want to use these studies to consider the relation between large N, macroscopic quantum systems, the uncertainty relation and classicality. We first give a brief account of how GP treated this problem, then point out the deficiencies in the criterion they used for classicality. We then describe the necessary requirements for classicality as represented by the behaviour of the Wigner function, namely that it \textit{decoheres }in the presence of an environment and it possesses \textit{correlations} between the canonical variables as in a classical trajectory in phase space.

GP considered an eternal inflation situation with the background metric in the form

\begin{equation}
\rmd s^2 = - \rmd t^2 + a (t)^2 \rmd x^2
\end{equation}
where the scale factor $a(t) = \rme^{Ht}$ for eternal inflation with the Hubble expansion rate given by $H(t) = \sqrt{8 \pi G \rho_0 /3}$
where $\rho_0$ is the vacuum energy density of the false vacuum.
The inflaton field $\phi$ dynamics is described by the action
\begin{equation}
S = \int \rmd^4x \sqrt {-g}[ -\ha g^{\mu\nu} \partial_\mu \phi \partial_\nu \phi - V (\phi)]
\end{equation}
with the potential $V(\phi)$ given by
\begin{equation}
V(\phi) = \frac{1}{4} \lambda \left( \phi^2 - \frac{\mu^2}{\lambda} \right)^2
\end{equation}
where $\mu$ is related to the value of the field $\phi_c = \frac{\mu}{\sqrt{\lambda}}$ at the minimum of the potential. The false vacuum energy is given by $\rho_0 = \frac{\mu^4}{4 \lambda}$, giving an expansion rate $H = (\frac{2\pi G}{3 \lambda})^{\ha} \mu^2 $.
From the action the inflaton field obeys the dynamical equation
\begin{equation}
\ddot \phi + 3H \dot \phi - a^{-2}(t) \Delta^{(3)} \phi = - \partial V / \partial \phi
\end{equation}
where  $\Delta^{(3)}$ denotes the 3-dim Laplace operator on the spatial hypersurface.

GP then expanded the free field (assumed to be confined within a 3-dim box of length $L$) in Fourier modes with $\sigma_{\kappa}$ being the amplitude for the $\pm$ propagating components of the $k$th mode and the zero mode (treated separately because of infrared divergence considerations) and its conjugate momentum $\pi_\kappa $. 

As initial condition GP assumed that at early times the system was in thermal equilibrium with a heat bath at temperature $T_0$ which rapidly drops as the universe inflates away $T = T_0 \rme^{-Ht}$. The expectation value  of physical quantities such as $\sigma_{\kappa}$ with respect to this thermal ensemble can be calculated by taking the ensemble average, 
\begin{equation}
<\sigma_\kappa^2(t)> = \ha (L/2\pi)^3 |\psi({\bf k}_{\kappa}t)|^2 \coth (\Theta_\kappa /2)
\end{equation}
where $\psi$ is given by the Hankel functions of the first kind and $\Theta_{\kappa}$ is the initial value of $\frac{\hbar \omega}{T(t)}$ at $t \rightarrow - \infty $.

To verify that the system behaves classically at late times GP  calculated  how the uncertainty function changes in time.
\begin{equation}
U_\kappa (t)= \frac{\left( <\sigma_{\kappa}^2><\pi_{\kappa}^2> \right)^\ha }{\hbar/2}
\end{equation}
The denominator $ \hbar / 2$ is the value for  minimal uncertainty in the case of a Gaussian wave packet at zero temperature.  Physically the $\kappa$ mode amplitude can be viewed as behaving classically when its effective wavelength $\lambda_{eff} = 2 \pi /k_{eff}$ (where $k_{eff} = a(t)/p(t)$, with $a(t)= \rme^{Ht}$, $p_{\kappa} = \sqrt{{\bf k}_{\kappa}^2 + \gamma^2}$, $\gamma = \sqrt{\lambda / \hbar} T_0 /2 $ ) 
is much larger than the horizon length.

Hu and Zhang \cite{HuZhaUnc} (see also Anastopoulos and Halliwell \cite{AnaHal}) have derived an expression for the uncertainty relation at finite temperature using the Wigner function of an oscillator interacting with a thermal bath modelled by $N$ harmonic oscillators as weighing function.  They also discussed how the contributions of quantum fluctuations weighted against thermal fluctuations and the criterion of classicality and thermality.

As later investigations showed this criterion used by GP for classicality is suggestive but incomplete.    Two conditions need be satisfied by the (reduced) Wigner function from the equivalent (reduced) density matrix (`reduced' refers to the situation where the system is coupled to an environment and rendered open). Wigner function has long been viewed as the quantum object `closest to' the classical distribution function. Up to the early 90's before  decoherence was considered in earnest it was viewed that if the Wigner function of a quantum system peaks along a classical trajectory, meaning that there is correlation between the coordinate x and momentum p, then one could view the system as classical. This criterion is shown to be valid only for Gaussian systems (such as a free harmonic oscillator with quadratic potential coupled bi-linearly to an environment of simple harmonic oscillators) because otherwise it may not even be positive definite \cite{Anderson}. In the Gaussian approximation the Wigner function is positive for all times but as was further pointed out \cite{HabLaf} it does not describe classical correlations
unless the system is coupled to an environment. Thus classical correlation and environment-induced decoherence are two necessary conditions for a quantum system to show classical behaviour.

When the wave packet  spreads out further away from the top of the potential the Gaussian approximation breaks down. As shown by Lombardo et al (LMM) \cite{LMM}, as the coupling between the system and the environment increases, its decoherence time decreases. Due to the nonlinearities of the
potential, when the coupling vanishes there is no classical limit, not even classical correlations. In fact, as reported by Antunes et al \cite{ALM}, in a quenched phase transition, even after classicalization has been reached, the system may display quantum behaviour again.

With these new understandings LMM reanalyzed GP's model with and without an environment, using the quantum Brownian motion \cite{FeyVer,CalLeg} model, especially results from  the non-Markovian master equation for a general environment \cite{HPZ}. They found that when the system is isolated, due to the high squeezing of the initial wave packet $x$ and $p$ become classically correlated. The density matrix is not diagonal. The correlation time depends on the shape of the potential. Only when the particle is coupled to an environment will a {\it bona fide} quantum to classical transition occur. The Wigner function becomes peaked around a classical trajectory and the density matrix diagonalizes. The decoherence time depends on the diffusion coefficient in the master equation.


Now that we have analyzed the quantum mechanical model in detail how would a macroscopic system as modelled by the O(N) quantum field theory behave?
Symmetry breaking in a $\phi^4 O(N)$ model has been studied in detail by many groups  \cite{CHKM97,Boy}. Lombardo et al \cite{LMM} carried out numerical analysis of this model and made the following observations: at late times the long-wavelength modes $k_l \ll t ^{-1}$  in the wave functional become classically
correlated. This is analogous to the situation described above for the inverted harmonic oscillator. For these modes the width of the Gaussian wave function increases linearly with time. Moreover, as the width of the wave function or the density matrix increases, the Wigner function becomes sharply peaked around the classical trajectory.  We see that in the large N limit, and at long times, the dynamical evolution of the O(N) model shows classical correlations but not true quantum to classical transition. Just as in  the inverted oscillator without environment, the density matrix does not become diagonal. The correlation time depends on the details of the potential.

In conclusion Lombardo et al \cite{LMM} observe that in order to get a Wigner function that is positive definite and peaked around a classical trajectory at long times, it is necessary to
have both vanishing effective mass and a Gaussian wave
function. Therefore, it is quite possible that in a field theory
calculation with finite N there will be no classical limit nor
classical correlations unless one allows the field to interact with an environment. On this case for an analysis of how the nonequilibrium evolution of a condensate and its fluctuations  vary with N, Baacke and Michalski \cite{BaaMic} have provided analytic expressions for general N and numerical solutions for N=1,4, 10. This should be compared with the analytic expressions for the effective potential calculated up to NLO for finite N given by Mihaila et al for an O(N) model, which we will discuss in section~\ref{subsec:HowlargeN}.

\subsection{Criteria of classicality: coarse-graining of quantum phase,  entropy generation}

We next use the entropy function derived from the number of particles and their quantum correlation in a quantum Vlasov equation to illustrate the notion of classicality. The large N aspect enters because quantum Vlasov equation is obtained as a mean field theory derivable as the leading order large N approximation from  scalar QED, where N is the number of identical copies of the charged matter field placed in an electromagnetic (EM) field . This is sometimes referred to as the semiclassical limit since the matter field is fully quantized and the EM field is treated classically.  Kluger, Mottola and Eisenberg \cite{KME} considered  the case of a spatially homogeneous electric field, represented by the vector potential in
the Coulomb gauge, $\mathbf{A}= A(t) {\hat {\mathbf{z}}},\,\,A_0 = 0 \,,$ whence the electric field is given by
$ \mathbf{E} = - \dot A {\hat {\mathbf{z}}} = E {\hat {\mathbf{z}}}\,. $

Assuming also that the field lives in a finite large volume $V$ we can expand the
charged scalar field operator in Fock space in Fourier modes. Since
particles are physically distinct from antiparticles, we need two
independent sets of destruction operators
\begin{eqnarray}
\Phi ({\mathbf {x}},t)&=&\frac{1}{\sqrt{V}}\sum_{\mathbf {k}}\rme^{\rmi \mathbf {k}\cdot \mathbf {x}%
}\varphi _{\mathbf {k}}(t)\\
&=&\frac{1}{\sqrt{V}}\sum_{\mathbf {k}}\left\{ \rme^{\rmi \mathbf {k}%
\cdot \mathbf {x}}f_{\mathbf {k}}(t)a_{\mathbf {k}}+\rme^{-\rmi\mathbf {k}\cdot \mathbf {x}}f_{-\mathbf {k}%
}^{*}(t)b_{\mathbf {k}}^{\dagger }\right\} \ .
\end{eqnarray}

Denote the time-independent annihilation operator of a particle in mode $%
\mathbf {k}$ by $a_{\mathbf {k}}$ and the creation of an anti-particle in mode $-%
\mathbf {k}$ by $b_{\mathbf {k}}^{\dagger }$. They obey the commutation relations
\begin{equation}
\lbrack a_{\mathbf {k}},a_{\mathbf {k}^{\prime }}^{\dagger }]=[b_{\mathbf {k}},b_{\mathbf {k}%
\mathbf{^{\prime }}}^{\dagger }]=\delta _{\mathbf {k}\mathbf {k}\mathbf{^{\prime }}%
}\,.  \label{aadag}
\end{equation}
Therefore $
N_{+}(\mathbf {k}) \equiv \langle a_{\mathbf {k}}^{\dagger }a_{\mathbf {k}}\rangle \, ,
N_{-}(\mathbf {k}) \equiv \langle b_{\mathbf {k}}^{\dagger }b_{\mathbf {k}}\rangle \, $
are the mean numbers of particles and antiparticles respectively. Without
loss of generality we can make use of the freedom in defining the initial
phases of the mode functions to set the correlation densities $\langle a_{%
\mathbf {k}}a_{\mathbf {k}}\rangle =\langle b_{\mathbf {k}}b_{\mathbf {k}}\rangle =0$. In a
Hamiltonian description we can take for each mode ${\mathbf {k}}$
\begin{equation}
\varphi _{\mathbf {k}}(t)\equiv f_{\mathbf {k}}(t)a_{\mathbf {k}}+f_{\mathbf {k}}^{*}(t)b_{-%
\mathbf {k}}^{\dagger }  \label{fcoor}
\end{equation}
as the (complex) generalized coordinates of the field $\Phi $ and
\begin{equation}
\pi _{\mathbf {k}}(t)=\dot{\varphi}_{\mathbf {k}}^{\dagger }(t)=\dot{f}_{\mathbf {k}%
}^{*}(t)a_{\mathbf {k}}^{\dagger }+\dot{f}_{\mathbf {k}}(t)b_{-\mathbf {k}}\,,
\label{fmom}
\end{equation}
as the momentum canonically conjugate to it. By virtue of the
commutation relation \eqn{aadag} they obey the canonical commutation
relation,
\begin{equation}
\lbrack \varphi _{\mathbf {k}},\pi _{\mathbf {k}\mathbf{^{\prime }}}]=\rmi \hbar \delta
_{\mathbf {k}\mathbf {k}\mathbf{^{\prime }}}\,,
\end{equation}
provided that the mode functions satisfy the Wronskian condition
\begin{equation}
\left( f_{k},f_{k}^{*}\right) =\rmi \hbar .  \label{wron}
\end{equation}

The complex amplitude function $f_{\mathbf {k}}(t)$ of the $\mathbf
{k}$ th mode satisfies the wave equation
\begin{equation}
\frac{\rmd^2 f_{\mathbf {k}}}{\rmd t^2}+\omega_{\mathbf {k}}^2(t) f_{\mathbf {k}}(t)=0,
\label{parameq}
\end{equation}
where the time dependent frequency $\omega _{\mathbf {k}}^{2}(t)$ is
given by
\begin{equation}
\omega _{\mathbf {k}}^{2}(t)=\left( \mathbf {k}-e\mathbf{A}\right)
^{2}+m^{2}=(k_{z}-eA(t))^{2}+k_{\perp }^{2}+m^{2}\,.  \label{plasmafreq}
\end{equation}
where $k_{z}$ is the constant canonical momentum in the $\mathbf{\hat{z}}$
direction while the physical (gauge-invariant) kinetic momentum is given by
\begin{equation}
p_{z}(t)=k_{z}-eA(t), \qquad \dot{p}_{z}=-e\dot{A}=eE  \label{cankin}
\end{equation}

For a spatially
homogeneous electric field ($i.e.,\mathbf {\nabla}\mathbf{\cdot E}=0$), by
Gauss' Law, the mean charge density must vanish, i.e,
$ j^{0}(t)=0. $ and the mean current in the $\mathbf{\hat{z}}$ direction is
\begin{equation}
j(t)=2e\int \,\rmd^{3}\mathbf{k}\;[k_{z}-eA(t)]|f_{{\mathbf {k}}}(t)|^{2}(1+N_{+}(\mathbf {k}%
)+N_{-}(-\mathbf {k}))  \label{curr}
\end{equation}
One can further restrict to the subspace of states for which
\begin{equation}
N_{+}(\mathbf {k})=N_{-}(-\mathbf {k})\equiv N_{\mathbf {k}}  \label{special}
\end{equation}
Clearly the vacuum $N_{+}(\mathbf {k})=N_{-}(-\mathbf {k})=0$ (as well as a thermal
state) belongs to this class of states.

Particle pairs will be produced in a strong background field, and in turn,
affect the strength and evolution of this background field. This first step is called a `test field' approximation where the background field is assumed fixed. The second step of including the effects of created particles is called a `backreaction problem'. It demands a self-consistent solution of the mean electric
field $\mathbf{E}(t)$ coupled to the expectation value of the current $j(t)$ of the quantum charged scalar field $\varphi _{\mathbf {k}}(t)$.
In a spatially homogeneous electric field, the only nontrivial Maxwell equation is simply,
\begin{equation}
-\dot{E}(t)=\ddot{A}(t)=j(t)  \label{max}
\end{equation}
where the current is given by \eqn{curr}. Since the charged scalar
field
depends on the vector potential $A$ to begin with, $f_{\mathbf {k}}(t)$ and $%
A(t) $ need be solved self-consistently from \eqn{parameq} with
\eqn {plasmafreq} and \eqn{max}.

\subsubsection{Particle production}

In a time dependent background the equation for the mode functions \eqn{parameq} would in general admit time dependent solutions. This means that one set of particle states which define a Fock space at one moment of time would change at a later time into another set of particle state in a different Fock space. In particular, an initial vacuum could evolve into an n-particle state later, which means particles are being produced. This is best expressed in terms of the Bogoliubov transformation \cite{Parker} between these two Fock space basis $a_{\mathbf {k}}$ and $\tilde{a}_{\mathbf {k}}$.  Consider for the moment just particles, no anti-particles, and no particle interactions (see \cite{HuKan,Elze} for that case). Let the first basis $a_{\mathbf {k}}$ be associated with modes $\left( f_{\mathbf {k}},f_{\mathbf {k}}^{*}\right) ,$ the second basis $\tilde{a}_{\mathbf {k}}$ with modes $%
\left( \tilde{f} _{\mathbf {k}}, \tilde{f}_{\mathbf {k}}^{*}\right).$
We may expand the field operators in either base, such as
\begin{equation}
\varphi _{\mathbf {k}}\left( t\right) =f_{\mathbf {k}}\left( t\right) a_{\mathbf {k}}+f_{%
\mathbf {k}}^{*}\left( t\right) a_{-\mathbf {k}}^{\dagger }  \label{d19c}
\end{equation}
in the first case, and
\begin{equation}
\varphi_{\mathbf {k}}\left( t\right) =\tilde{f}_{\mathbf {k}}\left( t\right) \tilde{a%
}_{\mathbf {k}}+\tilde{f}_{\mathbf {k}}^{\ast }\left( t\right) \tilde{a}_{-\mathbf {k}%
}^{\dagger }
\end{equation}
in the second. Since both sets of solutions of the mode equations are
complete, we must have
\begin{equation}
f_{\mathbf {k}}\left( t\right) =\alpha _{\mathbf {k}}\tilde{f}_{\mathbf {k}}\left(
t\right) +\beta _{k}\tilde{f}_{\mathbf {k}}^{*}\left( t\right) ,  \label{Bogf}
\end{equation}
and its inverse
\begin{equation}
\tilde{f}_{\mathbf {k}}(t)=\alpha _{\mathbf {k}}^{*}f_{\mathbf {k}}(t)-\beta _{\mathbf {k}%
}f_{\mathbf {k}}^{*}(t).  \label{Bogtf}
\end{equation}

The Wronskian condition $\left( f_{\mathbf {k}}, f_{\mathbf {k}}^{*}\right) =\left(
\tilde{f}_{\mathbf {k}},\tilde{f}_{\mathbf {k}}^{*}\right) = \rmi\hbar $ imposes a
condition on the Bogoliubov coefficients
\begin{equation}
|\alpha_{\mathbf {k}}|^{2}-|\beta_{\mathbf {k}}|^{2}=1  \label{can}
\end{equation}
for each $\mathbf {k}$. We can thus write
\begin{eqnarray}
|\alpha _{\mathbf {k}}(t)| &=&\cosh r_{\mathbf {k}}(t)\,,  \nonumber \\
|\beta _{\mathbf {k}}(t)| &=&\sinh r_{\mathbf {k}}(t)\,.  \label{magbog}
\end{eqnarray}
where $r_{\mathbf {k}}(t)$ is called the squeeze parameter for mode $\mathbf {k}$ a
terminology adopted from quantum optics. For a description of particle creation in the squeezed state language, see, e.g., Chapter 4 of \cite{CH08}.

The above only describes how one set of Fock states defined at one time is related to another set, but whether a vacuum can indeed be defined at any one time is a different and often more challenging problem.  We won't go into the details of this important issue, but just be satisfied for our present purpose with the fact that when the background varies sufficiently slowly one can invoke the concept of an adiabatic number state \footnote{This definition of a
number state makes use of the fact that under adiabatic evolution, particle
number is an adiabatic invariant, thus restricting its validity from the start
to extremely weak or slowing varying background fields. This level of approximation will not
give a good measure for on-going particle creation, as particle creation is
basically a nonadiabatic process. It is however useful for quantum kinetic
theory descriptions, where a quasi-particle approximation is usually
introduced which amounts to incorporating only the quantum radiative
corrections to the particles but not the fully field theoretical effects such as
particle creation.}. This  was used by \cite{KME} to calculate the charged particle production in an electric field  problem described above.

The adiabatic number state $\tilde{f}_{\mathbf {k}(0)}^{+}(t)$  corresponds to the $0 th$ order adiabatic vacuum [defined in (4.46) of \cite{CH08} as $f_{{\mathbf
{k}}(0)}^{+}$ ]
\begin{equation}
\tilde{f}_{\mathbf {k}}^{(0)}(t)\, \equiv
\sqrt{\frac{\hbar }{2\omega _{\mathbf {k}}(t)}}\exp \left( -\rmi\Theta
_{\mathbf {k}}^{(0)}\right) \,,  \label{adbmod}
\end{equation}
where $\Theta _{\mathbf {k}}^{(0)}\equiv \int^{t}\omega _{{\mathbf {k}}}(t^{\prime
})\rmd t^{\prime }\,,\label{actvar}$ is the $0$th order adiabatic phase. At
this level of accuracy one measures particle numbers at all times with
respect to the initial vacuum state at time $t_{0}$.
The adiabatic particle number is defined to be \cite{KME}
\begin{eqnarray}
\tilde{N}_{{\mathbf {k}}}(t) &\equiv &\langle \tilde{a}_{{\mathbf {k}}}^{\dagger }(t)%
\tilde{a}_{{\mathbf {k}}}(t)\rangle =\langle \tilde{b}_{{-\mathbf {k}}}^{\dagger }(t)%
\tilde{b}_{{-\mathbf {k}}}(t)\rangle =|\alpha _{{\mathbf {k}}}|^{2}\langle a_{{\mathbf {k%
}}}^{\dagger }a_{{\mathbf {k}}}\rangle +|\beta _{{\mathbf {k}}}|^{2}\langle b_{-\mathbf {%
k}}b_{-\mathbf {k}}^{\dagger }\rangle  \nonumber \\
&=&\left( 1+|\beta _{{\mathbf {k}}}|^{2}\right) N_{+}({\mathbf {k}})+|\beta _{{\mathbf {k%
}}}|^{2}\left( 1+N_{-}(-{\mathbf {k}})\right)  \nonumber \\
&=&|\beta _{{\mathbf {k}}}|^{2}+(1+2|\beta _{{\mathbf {k}}}|^{2})N_{\mathbf {k}}=N_{\mathbf {%
k}}+\left( 1+2N_{{\mathbf {k}}}\right) \,|\beta _{\mathbf {k}}(t)|^{2}
\label{adbpar}
\end{eqnarray}
where the last line is valid only if the number of positive and
negative charges are equal (cfr. \eqn{special}).
The amount of particle production at time $t$ in this basis is given by the
expectation value of the number operator $\tilde{a}^{\dagger }\tilde{a}$ at
time $t$ with respect to the vacuum state $|>_{0}$ defined at $t_{0}$, (not
the vacuum state $|>_{t}$ defined at $t$).

\subsubsection{Number and correlation}

In this  time-dependent particle number basis an equation can be obtained for the time rate of change of the number of particles created in each mode by differentiating \eqn{adbpar},
\begin{equation}
\frac{\rmd}{\rmd t}\tilde{N}_{{\mathbf {k}}}=2\left( 1+2N_{{\mathbf {k}}}\right) \mathrm{Re}%
\,(\beta _{{\mathbf {k}}}^{*}\dot{\beta}_{{\mathbf {k}}}).
\end{equation}

We need an expression for ${\dot{\beta}}_{\mathbf {k}}$ in terms of $\alpha_{\mathbf {k}}
,\beta_{\mathbf {k}} $ and $\Theta _{\mathbf {k}}^{(0)}(t)\equiv \int^{t}\omega _{{\mathbf {k}}%
}(t^{\prime })\rmd t^{\prime }$ (henceforth we will omit the superscript 0 on $\Theta _{\mathbf {k}}^{(0)}$). They are given by 
\begin{equation}
\dot{\alpha}_{{\mathbf {k}}}=\frac{\dot{\omega}_{{\mathbf {k}}}}{2\omega _{{\mathbf {k}}}%
}\beta _{{\mathbf {k}}}\exp (2\rmi\Theta _{{\mathbf {k}}}),\;\;\;\dot{\beta}_{{\mathbf {k}}%
}=\frac{\dot{\omega}_{{\mathbf {k}}}}{2\omega _{{\mathbf {k}}}}\alpha _{{\mathbf {k}}%
}\exp (-2\rmi\Theta _{{\mathbf {k}}})\,.  \label{abeom}
\end{equation}
and thus,
\begin{eqnarray}
\frac{\rmd}{\rmd t}\tilde{N}_{{\mathbf {k}}}&=&{\frac{\dot{\omega}_{{\mathbf {k}}}}{\omega _{{%
\mathbf {k}}}}}\left( 1+2N_{{\mathbf {k}}}\right) \,\mathrm{Re}\left\{ \alpha _{{%
\mathbf {k}}}\beta _{{\mathbf {k}}}^{*}\exp (-2\rmi\Theta _{{\mathbf {k}}})\right\} \nonumber \\
&=&{\frac{%
\dot{\omega}_{{\mathbf {k}}}}{\omega _{{\mathbf {k}}}}}\,\mathrm{Re}\left\{ \mathcal{%
C}_{{\mathbf {k}}}\exp (-2\rmi\Theta _{{\mathbf {k}}})\right\} \,,  \label{Neom}
\end{eqnarray}
where we have defined the time-dependent pair correlation function,
\begin{equation}
\mathcal{C}_{{\mathbf {k}}}(t) \equiv \langle \tilde a_{{\mathbf {k}}}(t) \tilde b_{-%
\mathbf {k}}(t)\rangle = \left( 1 + 2N_{{\mathbf {k}}}\right)\alpha_{{\mathbf {k}}}
\beta_{{\mathbf {k}}}^*\,.  \label{pcor}
\end{equation}

The pair correlation $\mathcal{C}_{{\mathbf {k}}}(t)$ is a very
rapidly varying function, since the time dependent phases on the
right side of \eqn{pcor} \emph{add} rather than cancel. The phases,
however, nearly cancel in the final combination of \eqn{Neom} to
render $\tilde{N}_{\mathbf{k}}$ a slowly varying function. The time
derivative of the pair correlation function is given by,
\begin{eqnarray}
\frac{\rmd}{\rmd t}\mathcal{C}_{{\mathbf {k}}}&=&{\frac{\dot{\omega}_{{\mathbf {k}}}}{2\omega
_{{\mathbf {k}}}}}\left( 1+2N_{{\mathbf {k}}}\right) \exp (2\rmi\Theta _{{\mathbf {k}}%
})\,\left( 1+2|\beta _{{\mathbf {k}}}|^{2}\right) \nonumber\\
&=&{\frac{\dot{\omega}_{{\mathbf {k}}%
}}{2\omega _{{\mathbf {k}}}}}\left( 1+2\tilde{N}_{{\mathbf {k}}}\right) \exp
(2\rmi\Theta _{{\mathbf {k}}})\,.  \label{cordt}
\end{eqnarray}

\subsubsection{Quantum Vlasov equation}

Let us return now to the two equations for the rates of change of the
particle number and the quantum correlations. Solving \eqn{cordt}
formally for $\mathcal{C}_{{\mathbf {k}}}$, assuming that $\mathcal{C}_{{\mathbf {k}}%
}$ vanishes at some $t=t_{0}$ which could be taken to $-\infty $, and
substituting into \eqn{Neom} we obtain
\begin{equation}
\frac{\rmd}{\rmd t}\tilde{N}_{{\mathbf {k}}}=\frac{\dot{\omega}_{{\mathbf {k}}}}{2\omega _{{%
\mathbf {k}}}}\int_{t_{0}}^{t}\,\rmd t^{\prime }\,\left\{ {\frac{\dot{\omega}_{{\mathbf {%
k}}}}{\omega _{{\mathbf {k}}}}}(t^{\prime })\left( 1+2\tilde{N}_{{\mathbf {k}}%
}(t^{\prime })\right) \cos \left[ 2\Theta _{{\mathbf {k}}}(t)-2\Theta _{{\mathbf {k}}%
}(t^{\prime })\right] \right\} \,,  \label{nonlpc}
\end{equation}
\eqn{nonlpc} gives the rate of particle creation in an arbitrary time-varying mean field which may be called a ``quantum Vlasov equation''. Note the appearance of the Bose enhancement factor $(1+2%
\tilde{N}_{{\mathbf {k}}})$ in \eqn{nonlpc} indicates that both
spontaneous and induced particle creation are present. One important
feature of \eqn{nonlpc} is that it is nonlocal in time, the
particle creation rate depending on the entire previous history of
the system. Thus particle creation in general is a
\textit{non-Markovian} process \cite {BirDav,Rau94,RauMue96,SRSBTP97}. Note that
the nonlocal form of \eqn{nonlpc} results from solving one variable
$\mathcal{C}$ in terms of the other $\tilde{N}$, each obeying a
Hamiltonian equation of motion. This is a general feature of coupled
subsystems.

\eqn{nonlpc} becomes exact in the limit in which the electric
field can be treated classically, i.e the limit in which real and
virtual photon emission is neglected, and there is no scattering.
Inclusion of scattering processes leads to collision terms on the
right side of \eqn{nonlpc} which are also nonlocal in general. This
nonlocality is essential to the quantum description in which phase
information is retained for all times. The phase oscillations in the
cosine term are a result of the quantum coherence between the created
pairs, which must be present in principle in any unitary evolution.
However, precisely because these phase oscillations are so rapid it
is clear that the integral in \eqn{nonlpc} receives most of its
contribution from $t^{\prime }$ close to $t$, which suggests that
some local approximation to the integral should be possible, provided
that we are not interested in resolving the short time structure or
measuring the phase coherence effects. The time scale for these
quantum phase coherence effects to wash out is the time scale of
several
oscillations of the phase factor $\Theta _{{\mathbf {k}}}(t)-\Theta _{{\mathbf {k}}%
}(t^{\prime })$, which is of order $\tau _{qu}={2\pi /\omega _{{\mathbf {k}}}}={%
2\pi \hbar /\epsilon _{{\mathbf {k}}}}$, where $\epsilon _{{\mathbf {k}}}$ is the
single particle energy.

\subsection{Density matrix and entropy}


After the elimination of the rapid variables ${\cal C}_{\bf k}$
defined in \eqn{pcor} in favour of the slow variables ${\cal N}_{\bf k}$
one can construct the density matrix in the adiabatic number basis
easily \cite{KME}. In a pure state  the only
nonvanishing matrix elements of $\rho$ are in uncharged pair states
with equal numbers of positive and negative charges, $\ell_{\bf k} =
n_{\bf k}^{(+)} = n_{\bf k}^{(-)}$, with $\ell_{\bf k}$ the number
of pairs in the mode $\bf k$, {\it viz.}
\begin{equation}
\langle 2\ell'_{\bf k} \vert \rho \vert 2\ell_{\bf k}\rangle
_{\rm pure} = \rme^{\rmi ({\ell_{\bf k}}^{\prime}- \ell_{\bf
k})\vartheta_{\bf k} (t) }\ {\rm sech}^2 r_{\bf k} (t)\
\left({\rm tanh}r_{\bf k} (t)\right) ^{{\ell_{\bf k}}^{\prime} +
\ell_{\bf k}} \label{adbden}
\end{equation}
where the magnitude of the Bogoliubov transformation, $r_{\bf k}(t)$
is defined in \eqn{magbog} and its phase, $\vartheta_{\bf k} (t)$
is determined by
\begin{equation}
\alpha_{\bf k}\beta_{\bf k}^{\ast} \rme^{-2i\Theta_{\bf k}} = - {\rm
sinh}r_{\bf k}\ {\rm cosh}r_{\bf k}\ \rme^{\rmi \vartheta_{\bf k}}\,.
\label{thetdef}
\end{equation}
Hence the off-diagonal matrix elements $\ell' \neq \ell$ of
$\rho$ are rapidly varying on the time scale $\tau_{qu}$ of the
quantum mode functions, while the diagonal matrix elements $\ell'
= \ell$ depend only on the adiabatic invariant average particle
number via
\begin{eqnarray}
\langle 2\ell_{\bf k} \vert \rho \vert 2\ell_{\bf k}\rangle _{\rm
pure} &\equiv& \rho_{2\ell_{\bf k}}={\rm sech}^2 r_{\bf k} {\rm
tanh}^{2\ell_{\bf k}} r_{\bf k} \nonumber \\
&=& \frac{\vert\beta_{\bf
k}\vert^{2\ell_{\bf k}} }{(1 + \vert\beta_{\bf
k}\vert^2)^{\ell_{\bf k}+1}} = \frac{{\cal N}_{\bf k}^{\ell_{\bf
k}}}{ (1 + {\cal N}_{\bf k})^{\ell_{\bf
k}+1}}\bigg\vert_{_{\rm pure}}\,, \label{rhodiag}
\end{eqnarray}
and are therefore much more slowly varying functions of time. The
average number of positively charged particles (or negatively
charged antiparticles) in this basis is given by
\begin{equation}
\sum_{\ell_{\bf k}=0}^{\infty}\ell_{\bf k} \rho_{2\ell_{\bf k}} =
{\cal N}_{\bf k}\,.
\end{equation}
Thus the diagonal and off-diagonal elements of the density matrix
in the adiabatic particle number basis stand in precisely the
same relationship to each other and contain the same information
as the particle number ${\cal N}_{\bf k}$ and pair correlation
${\cal C}_{\bf k}$ respectively.

From the (admittedly somewhat simplistic) perspective of the diagonization of the (reduced) density matrix as a signifier of classicality, this is similar to the criterion used in environment-induced decoherence alluded to in section 4.1, with one major difference: Here, quantum phase in the system is eliminated by choice of representation, while there, it is through the system's interaction with the environment. The former case needs physical justification for making such a choice, the latter needs explicit demonstration of how the pointer basis from measurement readings depend on the interaction Hamiltonian.  We shall continue the first route below.

\subsubsection{Classicality from ignoring the quantum phase}

In the density matrix \eqn{rhodiag} the diagonal elements
$\rho_{2\ell_{\bf k}}$ may be interpreted (for a pure state) as the
independent probabilities of creating $\ell_{\bf k}$ pairs of charged
particles with canonical momentum $\bf k$ from the vacuum. This
corresponds to disregarding the intricate quantum phase correlations
between the created pairs in the unitary Hamiltonian evolution. When
physics is expressed in the adiabatic particle number basis (the
Fock or N representation) the phase information is ignored. The quantum
density matrix in this representation produces an entropy function
which reflects the entropy associated with particle creation alone but says
nothing about  the evolution of the quantum phase or correlation. This
illustrates the crucial role played by the choice of representations
in the definition of entropy associated with particle creation
\cite{HuPav86}.

Results obtained from neglecting quantum phase are known to be quite
accurate for long intervals of time in the back-reaction of the
current on the electric field producing the pairs, because when
the current
is summed over all the $\bf k$ modes, the phase information in the
pair correlations cancels very efficiently. Thus for practical
purposes one can approximate the full Gaussian density matrix over
large time intervals by its diagonal elements only, in this basis.

\subsubsection{Entropy generation from particle creation}

Let us examine the reduced von Neumann entropy constructed from
the diagonal density matrix \eqn{rhodiag}
\begin{equation} S_{{\cal N}}(t) = -\sum_{\bf k}\sum_{\ell_{\bf k} =
0}^{\infty} \rho_{2\ell_{\bf k}} \ln \rho_{2\ell_{\bf k}}
\end{equation}
Upon substituting \eqn{rhodiag} into this, the sums over $\ell_{\bf
k}$ are geometric series which are easily performed. The von Neumann
entropy of this reduced density matrix
\begin{equation}
S_{{\cal N}}(t) = \sum_{\bf k} \left\{ (1 + {\cal
N}_{\bf k})\ln (1 + {\cal N}_{\bf k}) - {\cal N}_{\bf k}\ln {\cal
N}_{\bf k}\right\}
\end{equation}
is precisely equal to the Boltzmann entropy of the single particle
distribution function ${\cal N}_{\bf k}(t)$. Hence
\begin{equation}
\frac{\rmd }{\rmd t}S_{\cal N} = \sum_{\bf k} \ln \left( \frac{1 + {\cal
N}_{\bf k} }{{\cal N}_{\bf k}}\right) \frac{\rmd}{\rmd t}{\cal
N}_{\bf k} \label{entropy}
\end{equation}
increases if the mean particle number increases. This is always the
case {\it on average} for bosons if one starts with vacuum initial
conditions,  since $\vert \beta_{\bf k}\vert^2$ is necessarily
nonnegative and can only increase if it is zero initially
\cite{Kan88}. Locally, or once particles are present in the initial
state, particle number or the entropy \eqn{entropy} does not
necessarily increase monotonically in time.

Hence the notion of entropy associated with particle creation, and
the lore that it increases in time, is only valid for spontaneous
production of bosons from an initial vacuum state. This function
associated with fermions,  and that associated with stimulated
production of both boson and fermions, can decrease in time.

\subsection{Decoherence of hydrodynamic variables, conservation laws and quasi-classical domains}
\label{subsec:DecHyd}

To end this section on the criteria of classicality we want to describe an important formulation in how classicality emerges. This is  the  decoherent histories approach of Gell-Mann and Hartle \cite{GelHar1} applied to the quantum mechanics of closed systems, which is very different from the above-mentioned environment-induced decoherence scheme applied to open quantum systems. A central theme of this program is the decoherence of hydrodynamic variables, namely,  that the variables typically characterizing the quasiclassical domain of a  large and possibly
complex quantum system   that will become classical `habitually'  are the integrals over small volumes of locally conserved densities. Examples of local densities are number density, momentum density, energy density, charge density, which are collectively called hydrodynamic variables. The question of emergent classicality then consists first of understanding why these variables enjoy this special status, and second, of deriving the familiar hydrodynamic equations for these densities from decoherent histories. The representative references are \cite{GelHar2,HarLafMar,HalHydro,BruHal}. The treatment in Brun and Halliwell \cite{BruHal} contains examples for ease of understanding. For deeper conceptual issues related to quasi-classical domains read Dowker and Kent \cite{DowKen}.

We mention here those results which are directly relevant to our discussion of hydrodynamics, thermodynamics and classicality.

1) Following Gell-Mann and Hartle \cite{GelHar2} Hartle, Laflamme and Marolf \cite{HarLafMar} showed that exactly conserved quantities are exactly decoherent, and it is conceivable that the probabilities for such histories would peak about deterministic evolution equations. Brun and Halliwell \cite{BruHal} derived a formula which shows the explicit connection between local conservation and approximate decoherence.

2) Halliwell \cite{HalHydro} showed that a) for the case of the diffusion of the number density of a dilute concentration of foreign particles in a fluid, for certain physically reasonable initial states, the probabilities for the histories of the number density are strongly peaked about
evolution according to the diffusion equation.  b) When the initial state is a local equilibrium state, large-N statistics guarantees that the probabilities for histories of local densities are strongly peaked about a single history which describes the evolution of the  mean in the local equilibrium initial state.  This intuitively clear result readily connects with previous work, which shows that the mean values of local densities evolve according to hydrodynamic equations. The fact that there is essentially only one history
with nonzero probability means that there is approximate decoherence.

3) Brun and Halliwell \cite{BruHal} considered a class of models consisting of a large number of weakly interacting
components, in which the projections onto local densities may be decomposed into projections onto one
of two alternatives of the individual components. One model they gave is the placement of N particles into two partitions, the other model consists of a long chain of locally coupled spins for which the Hamiltonian conserves the total spin.  They computed the decoherence functional for histories of local densities, in the limit when the number of components is very large, they  find that decoherence requires two conditions: a) the smearing volumes must be sufficiently large to ensure approximate conservation, and b) the local densities must be partitioned into sufficiently large ranges to ensure
protection against quantum fluctuations.

The last two points are  of special interest to us: namely, 2b) referring to the evolution of the  mean in the local equilibrium initial state shows the relation of thermality and mean field dynamics. 3a) involving a large number of components N in a system and a large volume V smacks of thermodynamical limit, except that hydrodynamics ensures that conservation laws apply for these collective variables. Point 3b) above gives the condition for quantum fluctuations to be suppressed. We saw similar descriptions for the mean field and fluctuations in terms of leading order and NLO expansions for a large N quantum mechanical system.

\section{Next-to-leading order large N, quantum fluctuations}
\label{sec:NLOLN}

So far we have discussed the leading order theory of an O(N) model obtained from a large N expansion. This  mean field theory and its behaviour is usually referred to as ``classical". Assuming such a connection which we will examine critically in this section, we may ask: 
1) Can one identify a number $N_c$ whereby a system containing $N$ components with $N > N_c$ will begin to acquire classical features? 2) Since a classical system is devoid of quantum fluctuations, can one extract the information related to quantum fluctuations from higher order expansions in large N? Following this logic one may hope to come up with a systematic way to identify the demarkation zone between  classical and quantum features in terms of N for a quantum system with a large collection of components.

But as we forewarned in the Introduction, this equivalence of mean field with classical features is valid only for Gaussian theories. More accurately one should think about mean field rather than classical theory in the leading order large N  and deviations from the mean in the NLO large N. One should therefore ask, 3) Does one see a change of qualitative behaviour in the NLO or higher order large N expansion from the mean field theory?
What measures deviation from mean field? Perhaps one should not think of quantum fluctuations, but correlations, as in the BBGKY hierarchy? (The role of 2PIEA in closed (democratic) systems such as systems with N components will be discussed in our next paper.)  Finally, to see the difference of behaviour from  the O(N) model we have studied above one should also ask,  4) Would the large N limit of a gauge field theory behave differently, and how so? What can one say about deviation from the mean field and how the contribution of quantum fluctuations enter in these theories?

In the following three subsections we will address issues raised above: subsection~\ref{subsec:HowlargeN} in the context of O(N) theory for question  1), subsection~\ref{subsec:Grav} in the context of stochastic gravity for queries 2) and 3), and subsection~\ref{subsec:Gauge} in the context of large N gauge field theory for question 4).

\subsection{How large an N will NLO give sensible results? Quantum features beyond mean field theory?}
\label{subsec:HowlargeN}

For this question let us return to the results reported in section~\ref{sec:ONlargeN} pertaining to next-to-leading order (NLO) large N expansion in an O(N) model. The authors of  \cite{MACDH} show that the naive  large $N$
expansion violates unitarity (or more generally, positivity) leading
to an instability at least for $N$ less than some value $N_T$. They have provided
numerical evidence that a sharp threshold at $N\sim N_T$ exists, where for $N > N_T$ they saw no instability. They trace this behaviour
to be related to the nature of the effective potential at NLO: At this order, the effective potential has the
property of not being defined everywhere for values of $N < N_{c}$, where $N_{c}$ depends on the values of the parameters specifying the
potential, and $N_T\sim N_{c}$. However, for $N > N_{c}$, the effective potential is defined everywhere and there is no instability.

Probing further using the effective potential for the NLO expansion these authors asked the following question: ``How does the point $y_{\rm{min}}$,
below which the next-to-leading order large $N$ effective potential does not exist, changes as a function of $N$?" We know that at
``infinite $N$,'' (leading order), $y_{\rm{min}}=0$, but it is
important to know how this limit is reached. For instance, is there a
finite value of $N$ beyond which $y_{\rm{min}}=0$?  They plotted (figure 5 of  \cite{MACDH}) $y_{\rm{min}}$ as a function of
$N$ for the set of parameters chosen in their equation (7.15):  At the next-to-leading order large $N$ level, for $N \le 18.6$, $ y_{\rm{min}}$ is finite, but for $N \ge 18.6$, it hits the origin. Thus for $N \ge 18.6$, the authors concluded,  one can associate a quantum state (though not known explicitly) with the next-to-leading order approximation.

These findings by  \cite{MACDH} are interesting by implications. If one adheres to the conventional wisdom  based on the loose concept that mean field theory obtained from LO approximation describes classical physics then one might be tempted to think of  the NLO expansion as containing some quantum features. Does $N \ge N_c$ signals a transition from quantum to classical?

\subsubsection{Is there a critical $N_c$ which marks the appearance of MQP?}

If  $N=\infty$ is used as a criterion to view an N component quantum system as reaching its classical limit, it seems logical to ask, ``What is the smallest number $N_c$ for such classical behaviour in this  quantum system to disappear?  The related question for our concern is, ``Does the number $N_c$ obtained by these authors carry such a meaning, as a demarkation between classical and quantum behaviors?" If so this is a very neat result because it provides a quantitative measure in how `large' N needs to be to mark the appearance of MQP. We shall explain in what follows that unfortunately this is not the case. Life would have been much easier if it were. The key observation is that it is only for Gaussian theories that the mean field theory from LO LN approximation is equivalent to a classical theory. The critical $N_c$ is only a measure of the validity of the NLO approximation, not a measure of the importance of quantum contributions. Quantum features are pervasive at all levels of approximation.  We will explore this issue further in the context of critical phenomena in the sequel paper. \\

{\it Convexity and positivity.} Let us look at some technical features of the effective potential method which these authors used. The generating functional of connected Greens functions ($W \sim \ln Z$) is convex. This follows from the positivity of the two point function which is obtained by functionally differentiating $W$ twice. Legendre transformation takes a convex function to a concave function. Paying attention to the signs in the definitions it follows that the effective potential is also convex. As a result the convexity of the effective potential is a consequence (or a guarantee, depending on which way one looks at it) of the positivity.
This is the reason why the critical values $N_T$ and $N_c$ are about the same. Recall that $N_T$ is the critical value of $N$ below which an instability occurs (see figure 7 in MACDH for a plot of $<r^2>$ as a function of $t$) at finite time and $N_c$ is the critical value of $N$ below which the effective potential ceases to exist everywhere.

Perturbative expansions of the effective potential using the background field method run into difficulties for classical potentials (tree level) that are non-convex, which is the case described in MACDH. The loop expansion develops an imaginary part because of the unstable modes, which prompt tunneling or nucleation. If one were to sum up all the loop corrections the result would be a convex effective potential (with minima at the origin). However at finite order there usually are domains where the effective potential becomes imaginary.

The NLO LN expansion is not immune to this shortcoming. It sums up graphs from all orders in loop expansion but still the dynamics is altered. The same problem as in loop expansion occurs here as well. For $N > N_c$ convexity is restored.  The effective potential is convex for both classical and quantum O(N) models if we don't make any approximation. Hence the fact that convexity is restored for $N > N_c$ is a statement about the validity of the $1/N$ approximation, not about the system becoming classical or if enough quantum details have  been retained.


\subsubsection{Large N approximation does not capture full quantum dynamics}

As MACDH mention in their paper although NLO is a significant improvement over LO for sufficiently large N, at late times NLO fails to capture the nonlinear effects as well as LO does. Hence having $N > N_c$ (or $N > N_T$, since they are both of the same order) doesn't mean NLO approximation describes the dynamics accurately, it simply means the divergences and nonpositivity is avoided. At late enough times the approximation will fail (see figure 11 in MACDH).   In general to make a claim about the validity of an approximation time need be specified, since any approximation will fail after some time. To inquire about the turnover from quantum dynamics to classical dynamics we are referring to the full dynamics, not just for short time transient or long time asymptote behaviours.

Thus, we see that $N > N_c$ is a criterion for the validity of the NLO approximation, as these authors meant it to be,  not so much about the quantum versus classical behaviour of the N component system. As a side note: it is the NLO approximation that has this pathology not the LO. LO approximation most likely gets inaccurate and unreliable before the NLO approximation diverges. So LO approximation behaving well doesn't say anything about its accuracy. Same applies to NLO approximation for $N > N_c$.


\subsection{Semiclassical and stochastic gravity as LO and NLO large N theories}
\label{subsec:Grav}

We want to use these two theories  as an instructional example to two important themes in our program, 1) the autocratic open quantum system (Langevin) paradigm,  where gravity is treated preferentially as the system of special interest over the matter field regarded as its environment where only its coarse-grained (or averaged or integrated over) effects are accounted for. This is how the Einstein-Langevin equation (ELE) \cite{ELE} was first derived. 2) the democratic large N (Boltzmann) system. We shall use an N component matter field as source  to illustrate how the semiclassical Einstein (SCE) equation and the ELE can be derived as the leading order O (1) and next to leading order O(1/N) theories in the large N expansion respectively.

Semiclassical gravity refers to the theory where gravity (metric function $g_{ab}$) is treated classically  but matter as a quantum field $\hat \phi_j$, the vacuum expectation value (vev) of whose (renormalized) stress energy tensor serves as the source to the semiclassical Einstein  equation
\begin{equation}
G_{ab}[g] =\kappa \left\langle \hat{T}_{ab}[g] \right\rangle
_\mathrm{ren} . \label{sce}
\end{equation}
where the subscript $ren$ denotes renormalized.

Stochastic gravity \cite{stogra} includes in the source of the SCE equation an additional term $\xi_{ab}$ accounting for the fluctuations of the stress  tensor operator of the quantum matter field through the Einstein-Langevin equation:
\begin{equation}
G_{ab}[g] =\kappa (\left\langle \hat{T}_{ab}[g]\right\rangle + \xi_{ab}[g] )
_\mathrm{ren} . \label{ele}
\end{equation}
The  stochastic source $\xi _{ab}[g]$ at the level of Gaussian approximation is completely
characterized by its correlation function in terms of the noise
kernel $\mathcal{N}_{abcd}(x,y)$. The noise kernel is the symmetrized connected part of the two-point quantum correlation function of the stress tensor operator with respect to the state of the
matter fields. It describes their stress-energy fluctuations via 
\begin{equation}
\left\langle \xi _{ab}[g;x) \xi _{cd}[g;y) \right\rangle _{\xi } =
\mathcal{N}_{abcd}(x,y) \equiv \frac{1}{2}\left\langle
\left\{\hat{t}_{ab}[g;x), \hat{t}_{cd}[g;y) \right\}
\right\rangle , \label{noise}
\end{equation}
where $\hat{t}_{ab} \equiv
\hat{T}_{ab}-\langle\hat{T}_{ab}\rangle$ and $\langle \ldots
\rangle$ is the usual expectation value with respect to the
quantum state of the matter fields, whereas $\langle \ldots
\rangle_\xi$ denotes taking the average with respect to all
possible realizations of the stochastic source $\xi_{ab}$.

\subsubsection{Open system approach: gravity interacting with quantum matter fields}

In the open quantum system approach one begins with a closed system, described by a density matrix
$\rho (x,q;x',q',t)$, defines a system $S$ of special interest described, say, by the $x$ variables or fields and look for the overall coarse-grained effects of its environments $E$, described by the $q$ fields, on this system. The influence functional (IF) formalism \cite{FeyVer} offers a very elegant way of capturing these effects on a microphysics level, with full self-consistency. If we are only interested in how the state of the system is influenced by the overall effects, but not the precise states, of the environment(s), then the reduced density matrix $\rho_r(x,x',t)=\int~\rmd q~\rho (x,q;x',q,t)$ could provide the relevant information. (The subscript $r$ stands for
reduced.) It is propagated in time from $t_i$ by the propagator ${\cal J}_r$:

\be
\rho_r(x,x',t)
=\int\limits_{-\infty}^{+\infty}\rmd x_i\int\limits_{-\infty}^{+\infty}\rmd x'_i~
 {\cal J}_r(x,x',t~|~x_i,x'_i,t_i)~\rho_r(x_i,x'_i,t_i~)
\label{pathint}
\te

Assuming that the action
of the coupled system decomposes as $S=S_s[x]+S_e[q]+S_{int}[x,q]$,
and that the initial density matrix factorizes (i.e.,
takes the tensor product form),
$\rho (x,q;x',q',t_i)=\rho_s(x,x',t_i)\rho_e(q,q',t_i)$, the
propagator for the reduced density
matrix is given by

$$
{\cal J}_r(x,x',t|~x_i,x'_i,t_i)=
 \int_{x_i}^{x_f} Dx~ \int_{x'_i}^{x'_f} Dx '~\rme^{\rmi S_{eff}[x, x', t]}
$$
where
\be
S_{eff}[x, x', t] \equiv S_s[x]-S_s[x']+S_{IF}[x,x',t] \label{Jr}
\te
is the full effective action and $S_{IF}$ is the influence action.
The influence functional $\cal F$ is defined as~\footnote{For more details and further discussions on how the IF is applied in statistical field theory and applications, see, e.g., \cite{Banff}.}
\begin{eqnarray}\fl
&\;&{\cal F}[x, x', t] \;\;\equiv\;\; \rme^{\rmi S_{IF}[x,x',t]} \nonumber\\
\fl &\;&\;\;\;\;\; \equiv \int~\rmd q_f~\rmd q_i~ \rmd q_i'~
  \int_{q_i}^{q_f}Dq~\int_{q'_i}^{q_f}Dq'~
\rme^{\rmi (S_e[q]+S_{int}[x ,q]-S_e[q']-S_{int}[x ',q'])} \rho_e(q_i,q'_i,t_i).
\label{SIF}
\end{eqnarray}
$S_{IF}$ is typically complex; its real part ${\cal R}$, containing the dissipation kernel
$ {\bf D}$, contributes to the renormalization of $S_s$, and
yields the dissipative terms in the effective equations of motion.
The imaginary part ${\cal I}$,
containing the noise kernel ${\bf N} $, provides the information about
the fluctuations induced on the system through its coupling to the environment.

For a system  with a quadratic action such as a simple harmonic oscillator,  bilinearly coupled to a Gaussian environment (made up of e.g., non-interacting harmonic oscillators)  as in the quantum Brownian motion (QBM) problem treated profusely in the literature (see, e.g., \cite{CalLeg,HPZ}) the influence action takes the generic form :
\begin{equation}
S_\mathrm{IF}\left[x,x'\right]
= Z\cdot \Delta + \Delta \cdot {\bf H} \cdot \Sigma +
\frac{\rmi}{2} \Delta \cdot {\bf N} \cdot \Delta  ,
\end{equation}
where $\Delta = x'-x, \Sigma = \ha (x +x')$ and $\cdot \equiv \int_i^f \rmd t$. $Z$ is a current term marking the contribution of the initial conditions, and  the kernels ${\bf H,N}$ are related to the dissipation and noise kernels. In exact analogy with the above we can derive the  dynamics of the gravitational field as an open system under the influence of $N$ quantum matter fields viewed as its environment by calculating the influence action
$S_\mathrm{IF}$ defined as: (details are in  \cite{HRV}, Appendix C)
\begin{equation}
\rme^{\rmi  S_\mathrm{IF}[h,h']} = \prod_{j=1}^{N} \int \mathcal{D}\phi_j
\mathcal{D}\phi'_j
\rme^{\rmi  S_\mathrm{m}[\phi_j,h] -\rmi S_\mathrm{m}[\phi'_j,h']}
\rho[\phi_j^{(i)},\phi_j^{\prime\,(i)}] .
\end{equation}
Up to quadratic order in the metric perturbations $h_{ab}$ it is given by
\cite{martin99}
\begin{equation}
S_\mathrm{IF}\left[\Sigma_{ab},\Delta_{ab}\right]
= N \left( Z\cdot \Delta + \Delta \cdot ({\bf H}+ {\bf M}) \cdot \Sigma +
\frac{\rmi}{8} \Delta \cdot \mathcal{N} \cdot \Delta \right) ,
\end{equation}
where we have introduced the semisum and difference variables
$\Sigma_{ab} = \left(h_{ab} + h'_{ab}\right)/2$ and
$\Delta_{ab}=h'_{ab} - h_{ab}$, $Z^{ab}(x) = - \ha
\langle\hat{T}^{ab}[\hat{\phi},g;x) \rangle$ and the kernels ${\bf H}$,
are ${\bf M}$ are defined as follows: 
\begin{eqnarray}
H^{abcd}\left( x,y\right)  = &-&\frac{1}{4} \mathrm{Im}
\left\langle T^{*}\hat{T}^{ab}\left[ \hat{\phi},g;x\right)
\hat{T}^{cd}\left[ \hat{\phi},g;y\right) \right\rangle
\nonumber \\
&+&\frac{\rmi}{8}\left\langle \left[ \hat{T}^{ab}
\left[ \hat{\phi},g;x\right) ,\hat{T}^{cd}\left[
\hat{\phi},g;y\right) \right] \right\rangle
\label{Hkernel} \\
M^{abcd}(x,y) = &-&\frac{1}{2}\left( \frac{1}{\sqrt{-g(x)}}
\frac{\delta \left(\left\langle
\hat{T}^{ab}[\hat{\phi},g_{ab};x) \right\rangle
\right)}{\delta g_{cd}(y)} \right) ,  \label{Mkernel}
\end{eqnarray}
where the notation $T^*$ was employed to indicate that the
spacetime partial derivatives appearing in the time-ordered
operators also act on the theta function implementing the time
ordering. The functional derivative appearing on the right-hand
side of (\ref{Mkernel}) should be understood to account only
for the explicit dependence on the metric: the implicit dependence
through the field operator $\hat{\phi}[g]$ is excluded.

As is known the kernels ${\bf H}$ and ${\bf M}$ exhibit divergences that are cancelled by
renormalizing the gravitational coupling constant and the cosmological
constant in the bare gravitational action as well as the coupling
constants of the counterterms quadratic in the curvature. (We will not
need terms of higher order in the metric perturbations because they
give contributions to the connected part of the CTP generating
functional of higher order in $1/N$.)

\subsubsection{Closed system: semiclassical and stochastic gravity from large N expansion}

We now  give a closed system treatment for 
the effects of an N component quantum matter field $\hat \phi_j$ in a background spacetime with metric $g_{ab}$. The semiclassical Einstein and the Einstein-Langevin equations can be identified as originating from the leading order and the next-to-leading order large N expansions. Treating perturbative quantum gravity (in terms of quantized linear metric perturbations $\hat{h}_{ab}$) interacting with N conformally invariant scalar fields $\hat \phi_j$, Hartle and Horowitz \cite{HarHor} computed the effective action to leading order in large N  and showed that  the semiclassical Einstein equation can be interpreted as the equation governing the evolution of the expectation value of the metric to leading order [which for this theory is actually $O(1)$]. The sources of the SCE equation are given
by the expectation value of the stress tensor operator of the quantum matter fields.  In the same vein Roura and Verdaguer \cite{RouVerLN} demonstrated that  the next to leading order  [which in this theory is $O(1/N)$] \footnote{Note the ordering scheme used in \cite{RouVerLN,HRV} is different from the customary usage. Their leading order in the large $N$ limit refers to the lowest order in $1/N$ with a nonvanishing contribution. Hence,
as we will see,  the leading order for the source of the semiclassical Einstein equation, which is proportional to the
expectation value of the stress tensor operator, is $1/N^0$, or $O(1)$ -- this is called leading order large N in \cite{HarHor}, whereas the leading order for the quantum two-point correlation functions is $1/N$. In the counting scheme of \cite{HarHor} this would be the next-to-leading order which yields stochastic gravity. In the way how it is presented here this latter expansion is to be understood in a NLO sense.} contribution to the \textit{quantum} correlation functions in a large $N$ expansion is equivalent to the \textit{stochastic} correlation functions obtained in the context of stochastic semiclassical gravity.
Let us see how this comes about.

Consider metric perturbations around a globally hyperbolic background spacetime with metric $g_{ab}$ and scalar curvature $R$ interacting with $N$ minimally coupled
free scalar fields $\phi_j (j=1,..,N)$. The action for the combined system is the sum of the gravitational action $S_\mathrm{g}$ plus the action for the
matter fields $S_\mathrm{m}$. The gravitational action is given
by the usual Einstein-Hilbert term, the corresponding boundary
term (which should be included to have a well-defined variational
problem) and the usual counterterms
required to renormalize the divergences arising when functionally
integrating the matter fields:
\begin{equation}
S_\mathrm{g} = \frac{N}{2\bar{\kappa}} \int_\mathcal{M} \rmd^4x \sqrt{-\tilde{g}}
R\left(\tilde{g}\right) +\frac{N}{\bar{\kappa}}
\int_{\mathcal{S}=\partial\mathcal{M}}\rmd^3x
\sqrt{\tilde{g}_{\mathcal{S}}}K_a^a(\tilde{g}) +\
(\mathrm{counterterms}) ,
\end{equation}
where $\tilde{g}_{ab}=g_{ab}+h_{ab}$ is the perturbed metric, $g_{ab}$
is the background metric and the gravitational coupling constant
$\kappa = 8\pi / m_p^2$ was rescaled to $\bar{\kappa}/N$ so that the
product of the rescaled gravitational constant times the number of
fields remains constant in the limit $N \rightarrow \infty$. The
action for the matter fields is
\begin{equation}
S_\mathrm{m} = -\sum_{j=1}^N \int_\mathcal{M} \rmd^4x\sqrt{-\tilde{g}}\frac{1}{2}
\left( \tilde{g}^{ab} \nabla_a\phi_j \nabla_b\phi_j
+ m^2 \phi_j^2 \right) ,
\end{equation}
where $m$ is the mass of the scalar field. We will use the expansion in
$1/N$ to systemize the contributions. At the end one can always
substitute back the rescaled gravitational constant in terms of the physical one.

To make connection of the large N expansion with the theory of stochastic gravity one can use
the procedure of Calzetta, Roura and Verdaguer \cite{CRV} for the QBM problem to derive the
symmetrized quantum correlation function for the metric perturbations (see Appendix C of \cite{CRV}):
\begin{eqnarray}
\frac{1}{2}\left\langle \left\{ \hat{h}_{ab}(x), \hat{h}_{cd}(x')
\right\} \right\rangle
&=& \left\langle \Sigma_{ab}^{(0)}(x) \Sigma_{cd}^{(0)}(x') \right\rangle
_{\Sigma_{ab}^{(i)}, \Pi^{cd}_{(i)}}
\nonumber \\
&\;&+ \frac{\bar{\kappa}^2}{N} \left( G_\mathrm{ret} \cdot \mathcal{N}
\cdot (G_\mathrm{ret})^{T}
\right)_{abcd} (x,x')
\label{correlation}.
\end{eqnarray}
There are two separate contributions to the two-point correlation function: the first one, called \emph{intrinsic} fluctuations, is related to the dispersion of the initial state for the metric perturbations, whereas the second one, called \emph{induced} fluctuations  is proportional to the noise kernel and accounts for the fluctuations
induced by their interaction with the quantum matter fields as its environment.  Under the aforementioned conditions, the symmetrized quantum correlation function for the metric perturbations is equivalent to the stochastic correlation function obtained in stochastic semiclassical gravity by solving the
Einstein-Langevin equation.

The above expos\'e serves to illustrate the following points:

1) In a large N expansion semiclassical gravity (SCG) corresponds to a mean field theory obtainable from the leading order $O(1)$ expansion, and stochastic gravity (STG) theory corresponds to the next-to-leading $O(1/N)$ order expansion. In this regard SCG and STG share the same properties as the $O(N)$ theory. Note however that stochastic gravity used in this more restrictive sense assumes Gaussian noise in the quantum matter field environment and weak perturbations off a background metric which is a solution of the SCE equations. In the more general sense such as used in \cite{QGkin} there is no such restriction and the parallel with the $O(N)$ model ends at the Gaussian level.

2) Stochastic gravity includes the fluctuations of the stress tensor in the matter sector via the noise kernel. This in turn enters into the determination of the two-point quantum correlation function for the metric perturbations. Contributions of higher order in $1/N$ enter either by including the vertices for the metric perturbations or terms from the influence functional evaluated beyond the Gaussian approximation. (e.g., in figure 1 of \cite{HRV} the two diagrams shown   give contributions of order $1/N^2$ to the two-point quantum correlation function for the metric
perturbations.)

3) From these theories one can also see the relation between  stochastic correlations and quantum fluctuations (via the loop expansion order), as well as the relation between correlation order in the Schwinger-Dyson hierarchy and the large N order. We shall dwell on the correlation aspects of MQP with the use of 2PI effective action  in our sequel paper.

Note that it is by virtue of its Gaussian property that  SCG as a mean field theory can be viewed as semiclassical. For STG if one does not make the Gaussian assumption for the noise, the Feynman-Vernon identity which enables one to interpret the quantum fluctuations of the matter field (as environment) as a classical stochastic source no longer holds. In what follows we give an example of field theories which are non-Gaussian. We will see that even at the rudimentary level the identification of mean field theory as classical is no longer valid.

\subsection{Large N gauge field and string theory}
\label{subsec:Gauge}

\subsubsection{1/N expansion in gauge theories}

't Hooft \cite{Hooft} introduced the $1/N$ expansion for nonabeliean gauge  theories in order to express physical quantities as a systematic expression in powers of $1/N$.
Consider the gauge group $SU(N)$. The gauge vector fields are Hermitian matrices $A_{\mu j}^i$, where both indices $i,j$ run from 1 to $N$. The field strength is
\begin{eqnarray}
F_{\mu \nu \ j}^{\ \ i} = \partial_{\mu}A_{\nu j}^{\ i} - \partial_{\nu}A_{\mu j}^{\ i} + i g [A_{\mu}, A_{\nu}]^i_j,
\end{eqnarray}
where $g$ is the gauge coupling constant. The Lagrangian density for a spinor particle of mass $m_{\psi}$ with wave function $\psi$ interacting with this gauge field is given by
\begin{eqnarray}
{\cal L} &=& - \frac{1}{4} {\rm Tr}(F_{\mu\nu}F^{\mu\nu}) - \bar{\psi}(i\gamma D + m_{\psi})\psi \nn \\
&=& - {\rm Tr}\Big( \frac{1}{2}(\partial_{\mu}A_{\nu})^2 - \frac{1}{2}\partial_{\mu}A_{\nu}\partial_{\nu}A_{\mu} +i g \partial_{\mu}A_{\nu}[A_{\mu}, A_{\nu}]
- \frac{1}{4} g^2 [A_{\mu}, A_{\nu}]^2 \Big) \nn  \\
&& - \bar{\psi}(i\gamma^{\mu} \partial_{\mu} + i g \gamma^{\mu}A_{\mu} + m_{\psi})\psi.
\end{eqnarray}
It is convenient to introduce a double line notation to keep track of the group indices i and j.
Whenever there is a close index loop in the Feynman diagram, we get a factor of $N$ from counting the values of the index can assume. This is how the dimension $N$ of the gauge group  enters. Since all three-point vertices  come with a factor of $g$,
and all four-point vertices  come with a factor of $g^2$, the $g$ and $N$ dependence of the amplitude for a general Feynman diagram
is seen to be $(g^2 N)^{F + 2 \chi -2} N^{2 - 2 \chi }$, where $F$ is the number of loops and $\chi$ is the genus, or number of ``holes"
of the polyhedron formed by the Feynman diagram. In the 't Hooft large $N$ limit while keeping $g^2 N \equiv \lambda$ fixed, we see that the $N$
dependence of the diagram is $N^{2 - 2 \chi}$. So the diagram without holes ($\chi=0$) dominate. In other words, in this limit, only the planar diagrams contribute.
The interactions between hadrons is expected to be a $O(1/N)$ effect. Hence in the large $N$ limit one can consider the problem of confinement and hadron mass
spectrum without the complication of residual hadron interactions.
The hope was that one could solve the theory with $N=\infty$ exactly followed by an expansion in $1/N=1/3$ for QCD.

Witten \cite{Witten79} referred to this limit as analogous to the classical limit for quantum mechanics, in the sense that for large $N$ there exists one single ``classical" gauge field, called ``master field",  which saturates the path integral and gives the dominant contribution. Such a classical field configuration was called a
Witten's argument was based on the assumption that all possible invariant operators satisfy the factorization property [which holds in any $U(N)$ symmetric theory of $N\times N$ matrix fields $A_{ij}(x)$]
\begin{eqnarray}
\langle f_1(A)...f_k(A) \rangle = \prod_{i=1}^k \langle f_i(A)  \rangle [1+ O(1/N)],
\end{eqnarray}
where the $f_i(A)$ are $U(N)$ symmetric operators of the form $f(A)= {\rm Tr}[A(x_1)...A(x_l)]$.

However, Haan \cite{Haan}  pointed out that there is a gap in Witten's argument for the existence of a master field in the large $N$ limit of QCD. He showed that the master field does not exist for the two-matrix model and the large $N$ limit is not a classical limit. Instead, the factorization of expectation values of invariant operators is shown to be a property analogous to the vanishing of fluctuations for macroscopic observables in the thermodynamic limit.

\subsubsection{Large N field theories and string theory}

At large $N$, QCD is expected to behave like a string theory. This is supported by the following observations. First, Feynman diagrams are organized in a topology expansion, just like string worldsheets. Second, the existence of Regge trajectories indicates a string-like behaviour. Third, QCD contains string-like objects, which are the electric flux tubes between quarks and antiquarks.  Maldacena's AdS/CFT correspondence \cite{Maldacena} conjectures  the equivalence of $N=4$ $SU(N)$ super Yang-Mills theory with gauge coupling $g_{YM}$ in 4 dimensions and Type IIB superstring theory with string coupling $g_s$ in $AdS_5 \times S^5$ space where both $AdS_5$ and $S^5$ have the same radius $L$. The parameters between these two theories are related by
\begin{eqnarray}
g_s = g_{YM}^2, \quad L^4=4 \pi g_s N (\alpha^{\prime})^2,
\end{eqnarray}
where $\alpha^{\prime}$ is related to the string tension. This conjecture is assumed to be valid for all values of $N$ and of $g_s = g_{YM}^2$.
Assuming the validity of Maldacena's conjecture, one can study different limits of field theory and string theory.

\begin{itemize}

\item
First take the 't Hooft limit. This is the case where $\lambda\equiv g_{YM}^2 N = g_s N$ fixed and letting $N \rightarrow \infty$.
In the Yang-Mills theory, this limit is well-defined, at least perturbatively,  and corresponds to a topological expansion of the field theory's Feynman diagrams. On the $AdS$ side,
the string coupling is $g_s=\lambda/N$ and the 't Hooft limit corresponds to weak coupling string perturbation theory. The $1/N$ expansion in gauge theory  corresponds to the $g_s$ string loop expansion in the string theory side. Hence the Maldacena conjecture in this case becomes a correspondence between large $N$ limit of gauge theories and the classical string theory. (Here classical means there is no string loop expansion in the $N\rightarrow \infty$ limit).

\item The large $\lambda$ limit. In this limit, one can expand the effective action in the $AdS$ side in terms of $\alpha^{\prime}$.
The distance scale is set by the $AdS$ radius $L$ which is related to the scale of Riemann tensor by $R\sim 1/L^2 = (g_s N)^{-\frac{1}{2}}/\alpha^{\prime}
= \lambda^{-\frac{1}{2}}/\alpha^{\prime}$. Hence in the large $\lambda$ limit, the classical Type IIB string theory on $AdS_5 \times S^5$ becomes classical Type IIB supergravity on $AdS_5 \times S^5$ and the $\alpha^{\prime}$ expansion in string theory corresponds to $\lambda^{-\frac{1}{2}}$ expansion in field theory.
(Here  gravity is weak because the curvature scale approaches zero in this limit and one can neglect the stringy correction and approximate it by the classical supergravity action).

\end{itemize}

To summarize, let us see what all this means for the issues we wish to address in this paper: classicality and large N from the perspective of string/gauge/gravity theories.

Old school field theorists often refer to the large $N$ limit of a Yang-Mills gauge theory as classical. Or stretching it a little, one may habitually  view all large $N$ theories as classical.   The advent of string theory alters fundamentally the way we look at nature, not just pertaining to Planck scale microphysics but also in relation to gauge fields and gravity, two essential elements of nature as we see it today.
To begin with, N enters in a new coupling constant called the 't Hooft coupling, in addition to the string tension,  the magnitude of both and their interplay alter a theory's qualitative behaviour in a fascinating way.
Another important new concept introduced by string theory is duality: between string theory and gauge theory, and between gauge theory and gravity.

The correspondence between gauge theory and string theory provides another window to examine the large $N$ behavior of gauge theory. Under this duality relation, a Yang-Mills gauge theory ingrains also the effects of gravitational interaction: Under different regimes of 't Hooft coupling, the behavior of large $N$ limit seen from the $AdS$ side of the correspondence can appear either as the suppression of string loop expansion and hence a classical string theory or the suppression of the stringy effect giving rise to a weakly interacting gravity theory where the quantum effect is negligible.

We can only give an inkling of the immense body of work accumulated in the last half a century for large N, especially  since 1998 when the AdS/CFT correspondence was proposed.  The simple message we wish to convey here is: ``exercise caution when tackling the meaning or finding the criteria of classicality": the explanation of the emergence of classical behavior or even the apparently straightforward identification of a classical regime from an interacting quantum theory is a subtle issue, as it may contain multiple structures and carry different dynamics in different interaction regimes.

\section{Conclusion}
\label{sec:conclusion}

Macroscopic quantum phenomena is a relatively new research venue, with exciting experiments and stimulating proposals. Surprisingly there are very little systematic theoretical investigations or even serious attempts we are aware of to construct a viable theoretical framework for this new endeavour.~\footnote{Maybe this is a sociological phenomenon, that exciting experiments are carried out by atomic-optical physicists while those who care or know more about the theories are mainly in the particle and gravitational physics community. If so this series of papers can be viewed as a modest attempt to  bridge that divide.} In this paper we ask only one simple question: Is it really that surprising that quantum features can appear in macroscopic objects? Note the surprise comes from the conventional and hitherto unchallenged view that macroscopic means classical. The purpose of our series of papers is to remove this degeneracy. By examining many well-known examples where theoretical analysis has been performed in detail for different systems, we find that there is no a priori good reason why quantum phenomena in macroscopic objects cannot exist. The challenge for theorists is to identify for various systems the physical conditions MQP manifest and in what ways.

In this paper we take the large N perspective, and try to understand what the leading order  and next-to-leading order large N expansions bring us. In general we can only say that the LOLN gives mean field theory and the NLO gives deviations from the mean field, both are still very much quantum in nature. Only for Gaussian models such as the $O(N)$ theory (zero-dimensional field theory, referring to the quantum mechanics of N oscillators) or in theories where the Gaussian approximation is imposed, such as semiclassical gravity and stochastic gravity, do we find that the mean field at LOLN is equivalent to classical, and the NLO incorporates deviations from the mean, which, only when one identifies the mean with classical would it be conjured as containing ``quantum fluctuations". By combing through these examples we are trying to change the common attitude on a very subtle point, namely,  ``Don't begin with classical and try to go backwards to quantum. One should always begin with quantum and only under very special conditions and  stringent criteria would a quantum system  begin to show certain classical features." We want to shift the burden of proof to those who believe they live in a classical world, to find these conditions and check every criteria. If they are not completely fulfilled, well then, perhaps one should just accept that the world is fundamentally quantum and it is only under specific yet commonly encountered conditions~\footnote{Specific and common are not contradictory conditions: equilibrium state is an example. It is one specific state out of a large number of states accessible to a large system,  but it is also the most commonly encountered --  with high probability -- when the system is placed in a heat bath. Behind this is the working of central limit theorem, quantum regression theorem in counting the statistics, or the conservation laws governing the hydrodynmic variable, as far as decoherence is concerned. These theorems and laws give a deeper meaning of `habitual'.} that it appears classical.

This message is of course not that foreign to researchers working in the last two decades on the quantum  to classical transition or the Q-C correspondence issues.  The half a dozen or so criteria we reviewed in this paper stem from  these investigations. What we have done somewhat different here is to examine every criteria anew with a macroscopic system, here with the aid of large N expansions. There is an  abundance of information in the particle physics literature on this subject. We brought forth some here to serve as illustrations for the one question we raised, but there is plenty more material which can be used fruitfully for performing quantitative analysis of MQP.

There are other important issues which we have left out. For example, here we use the number of components N as a measure of macroscopic, but it could be physically more  meaningful to incorporate interaction strengths and dynamics, in the renormalization group sense. Thus it could be  the ultraviolet versus the infrared behaviour of a (quantum) system which conveys a stronger microscopic versus macroscopic sense, as manifests in critical phenomena. It is near the critical point that the system's IR behaviour dominates, where the correlation function extends to infinity. This aspect of correlation, both classical and quantum, is the focus of our sequel paper. For interacting  quantum systems we will invoke the nPI effective action and the Schwinger-Dyson equations (corresponding classically to the BBGKY correlation hierarchy) to analyze what quantum versus classical means for large/small and/or strongly/weakly correlated systems. In that context we will include critical phenomena and other statistical mechanics considerations to examine this new issue of MQP at the foundation of quantum mechanics.



\ack{We learned from conversations or collaborations with, or reading the papers of, Charis Anastopoulos, Esteban Calzetta, Fred Cooper, Jonathan Halliwell, Albert Roura and Rafael Sorkin, and would be interested in how they would view and tackle the issues posed here. BLH wishes to thank the organizers of DICE2010, especially  Professor Thomas Elze, for their kind invitation. He also enjoyed the warm hospitality of the National Center for Theoretical Sciences (South) and the Department of Physics of National Cheng-Kung University, Tainan, Taiwan while part of this work was done. This work is supported by the National Science Council of Taiwan under grant NSC 97-2112-M-006-004-MY3 and by NSF grants PHY-0601550 and PHY-0801368 to the University of Maryland.}


\section*{References}

\begin{thebibliography}{99}

\bibitem{QCCDrexel} Feng D H and Hu B L (eds) 1997 {\it Quantum classical correspondence}
(Boston: International Press)




\bibitem{HuZhaUnc} Hu B L and Zhang Yuhong 1993 {\it Mod. Phys. Lett.} A {\bf 8} 3575;
{\it do.} 1995 {\it Int. J. Mod. Phys.} A {\bf 10} 4537

%

\bibitem{envdec}Paz J P and Zurek W 1999 Environment-induced decoherence and the transition from
quantum to classical {\it Lectures at the 72nd Les Houches Summer School on ``Coherent
Matter Waves"} {\it Preprint} quant-ph/0010011

Zurek W H 2003 {\it Rev. Mod. Phys.} {\bf 75} 715-75


\bibitem{nem} Armour A D, Blencowe M P and Schwab K C 2002 {\it Phys. Rev. Lett.} {\bf 88} 148301

Blencowe M P 2004 {\it Phys. Rep.} {\bf 395} 159-222


\bibitem{Arndt} Arndt M, Nairz O, Vos-Andreae J, Keller C, van der Zouw G and Zeilinger A 1999 {\it Nature} {\bf 401} 680



\bibitem{Marshall} Marshall W, Simon C, Penrose R and Bouwmeester D 2003 {\it Phys. Rev. Lett.} {\bf 91} 130401



\bibitem{EntSQL} Mueller-Ebhardt H, Rehbein H, Schnabel R, Danzmann K and Chen Y 2008 {\it Phys. Rev. Lett.} {\bf 100} 013601



\bibitem{Dorfman} Dorfman R 1999 {\it An Introduction to Chaos in Nonequilibrium Statistical Mechanics} (Cambridge: Cambridge Univ. Press)



\bibitem{Gaspard} Gaspard P 1998 {\it Scattering and Statistical Mechanics} (Cambridge: Cambridge Univ. Press)



\bibitem{QIPsi} Hu B L and Yu T (guest eds)  2009 {\it Quantum Information Processing} vol 8 Special Issue on Quantum Decoherence and Entanglement (Springer)



\bibitem{DVcriteria} DiVincenzo D P 2000 {\it Preprint} quant-ph/0002077



\bibitem{CHY} Chou C H, Hu B L and Yu Ting 2008 {\it Physica} A {\bf 387} 432-44


\bibitem{CHKMPA} Cooper F, Habib S, Kluger Y, Mottola E, Paz J P and Anderson P 1994 {\it Phys. Rev.} D {\bf 50} 2848




\bibitem{CHKM97} Cooper F, Habib S, Kluger Y and Mottola E 1997 {\it Phys. Rev.} D {\bf 55} 6471


\bibitem{HuRam97} Ramsey S and Hu B L 1997 {\it Phys. Rev.} D {\bf 56} 661


\bibitem{stogra} Hu B L and Verdaguer E 2008 {\it Living\ Rev.\ Rel.}\ {\bf 11} 3;
 {\it do.} 2003 {\it Class. Quant. Grav.} {\bf 20} R1




\bibitem{CJT} Cornwall J M, Jackiw R and Tomboulis E 1974 {\it Phys. Rev.} D {\bf 10} 2428



\bibitem{CH03} Calzetta E A and Hu B L 2003 {\it Phys. Rev.} D {\bf 68} 065027


\bibitem{HarHor} Hartle J B and Horowitz G 1981 {\it Phys. Rev.} D {\bf 24} 257






\bibitem{RouVerLN}  Roura A and Verdaguer E, unpublished. See also talk given by A.~Roura in Tainan, Taiwan, Jan 2007


%


\bibitem{MACDH} Mihaila B, Athan T, Cooper F, Dawson J and Habib S 2000 {\it Phys. Rev.} D {\bf 62} 125015



\bibitem{MDC} Mihaila B, Dawson J and Cooper F 2001 {\it Phys. Rev.} D {\bf 63} 096003



\bibitem{CHthermalize} Calzetta E and Hu B L 2002 Thermalization of an Interacting Quantum Field in the CTP-2PI Next-to-leading-order Large N Scheme {\it Preprint} hep-ph/0205271


\bibitem{Hooft} 't Hooft G 1974 {\it Nucl. Phys.} B {\bf 72} 461



\bibitem{Witten79} Witten E 1979 {\it Nucl. Phys.} B {\bf 160} 57




\bibitem{Maldacena} Maldacena J 1998 {\it Ad. Theor. Math. Phys.} {\bf 2 } 231


\bibitem{GuthPi} Guth A H and Pi S Y 1985 {\it Phys. Rev.} D {\bf 32} 1899


\bibitem{KME} Kluger Y, Mottola E and Eisenberg J M 1998 {\it  Phys. Rev.} D {\bf 58} 125015



\bibitem{HabibPRL}Habib S, Kluger Y, Mottola E and Paz J 1996 {\it  Phys. Rev. Lett.} {\bf 76} 4660




\bibitem{Habib08} Habib S 2004 Gaussian dynamics is classical dynamics {\it  Preprint} quant-ph/0406011


\bibitem{Coleman} Coleman S 1985 {\it Aspects of Symmetry} (Cambridge: Cambridge Univ. Press)





\bibitem{Haan} Haan O 1981 {\it Phys. Lett.} B {\bf 106} 207


\bibitem{Yaffe} Yaffe L 1982 {\it Rev. Mod. Phys.} {\bf 54} 407





\bibitem{ZHP} Zurek W H, Habib S and Paz J P 1993 {\it Phys. Rev. Lett.} {\bf 70} 1187





\bibitem{AnaHal} Anastopoulos C and Halliwell J J 1995 {\it Phys. Rev.} D {\bf 51} 6870




\bibitem{BlazotRipka} Blaizot J-P and Ripka G 1986 {\it Quantum Theory of Finite Systems}
(Cambridge MA: MIT Press) p~156



\bibitem{CJP} Coleman S, Jackiw R and Politzer H D 1974 {\it Phys. Rev.} D {\bf 10} 2491







\bibitem{Root} Root R 1974 {\it Phys. Rev.} D {\bf 10} 3322





\bibitem{Boy} Boyanovsky D, de Vega H J, Holman R, Lee D -S and Singh A 1995 {\it Phys. Rev.} D {\bf 51} 4419



\bibitem{CooperPi86} Cooper F, Pi S Y and Stancioff P 1986 {\it Phys. Rev.} D {\bf 34} 3831




%


%


\bibitem{Anderson} Anderson A 1990 {\it Phys. Rev.} D {\bf 42} 585



\bibitem{HabLaf} Habib S and Laflamme R 1990 {\it Phys. Rev.} D {\bf 42} 4056



\bibitem{LMM} Lombardo F C, Mazzitelli F D and Monteoliva D 2000 {\it Phys. Rev.} D {\bf 62} 045016




\bibitem{ALM} Antunes N D, Lombardo F C and Monteoliva D 2001 {\it Phys. Rev.} E {\bf 64} 066118





\bibitem{FeyVer} Feynman R and Vernon F 1963 {\it Ann. Phys.} (N.Y.) {\bf 24} 118



%

\bibitem{CalLeg} Caldeira A O and Leggett A J 1983 {\it Physica} {\bf121A} 587


%



\bibitem{HPZ} Hu B L, Paz J P and Zhang Y 1992 {\it Phys. Rev. D} {\bf 45} 2843;
{\it do.} 1993 {\it Phys. Rev. D} {\bf 47} 1576






\bibitem{BaaMic} Baacke J and Michalski S 2002 {\it Phys. Rev.} D {\bf 65} 065019


\bibitem{Parker} Parker L 1969 {\it Phys. Rev.} {\bf 183} 1057



\bibitem{HuKan} Hu B L and Kandrup H E 1987 {\it Phys. Rev.} D {\bf 35} 1776


\bibitem{Elze} Elze H-T 1996 {\it Phys. Lett.} B {\bf 369} 295


\bibitem{CH08} Calzetta E and Hu B L 2008 {\it Nonequilibrium Quantum Field Theory} (Cambridge: Cambridge Univ. Press)



\bibitem{BirDav} Birrell N D and Davies P C W 1982 {\it Quantum Fields in Curved Spaces} (Cambridge: Cambridge Univ. Press)




\bibitem{Rau94} Rau J 1994 {\it Phys. Rev.} D {\bf 50} 6911



\bibitem{RauMue96} Rau J and Mueller B 1996 {\it Phys. Rep.} {\bf 272} 1






\bibitem{SRSBTP97} Smolyansky S A, Roepke G, Schmidt S,  Blaschke D, Toneev V D and  Prozorkevich A V 1997 {\it Preprint} hep-ph/9712377







\bibitem{HuPav86} Hu B L and Pavon D 1986 {\it Phys. Lett.} B {\bf 180} 329



\bibitem{Kan88} Kandrup H E 1988 {\it Phys. Rev.} D {\bf 37} 3505

Kandrup H E 1988 {\it Phys. Rev.} D {\bf 38} 1773




%

\bibitem{GelHar1} Gell-Mann M and Hartle J B 1990 Quantum mechanics in the light of quantum
cosmology {\it Complexity, Entropy and the Physics of Information} ed W H Zurek (Reading, MA: Addison-Wesley)


%




\bibitem{GelHar2} Gell-Mann M and Hartle J B 1993 {\it Phys. Rev.} D {\bf 47} 3345



\bibitem{HarLafMar} Hartle J B, Laflamme R and Marolf D 1995 {\it Phys. Rev.} D {\bf 51} 7007




\bibitem{HalHydro} Halliwell J J 1998 {\it Phys. Rev.} D {\bf 58}
105015



\bibitem{BruHal} Brun T and Halliwell J J 1996 {\it Phys. Rev.} D {\bf 54} 2899



\bibitem{DowKen} Dowker F and Kent A 1996 {\it J. Stat. Phys.} {\bf 82} 1575








\bibitem{ELE} Calzetta E and Hu B L 1994 {\it Phys. Rev.} D {\bf 49} 6636

Hu B L and Matacz A 1995 {\it Phys. Rev.} D {\bf 51} 1577

Hu B L and Sinha S 1995 {\it Phys.  Rev.} D {\bf 51} 1587

Campos A and Verdaguer E 1996 {\it Phys. Rev.} D {\bf 53} 1927

Lombardo F C and Mazzitelli F D 1997 {\it Phys. Rev.} D {\bf 55} 3889


%

\bibitem{Banff} Hu B L 1994 Quantum statistical fields in gravitation and cosmology {\it Proc. Third
International Workshop on Thermal Field Theory and Applications} ed R Kobes and G Kunstatter  (Singapore: World Scientific) {\it Preprint} gr-qc/9403061


%


\bibitem{HRV} Hu B L, Roura A and Verdaguer E 2004 {\it Phys. Rev.} D {\bf 70} 044002


\bibitem{martin99} Martin R and Verdaguer E 1999 {\it Phys. Rev.} D {\bf 60} 084008

Martin R and Verdaguer E 1999 {\it Int. J. Theor. Phys.} {\bf 38} 3049



\bibitem{CRV} Calzetta E, Roura A and Verdaguer E 2003 {\it Physica} A {\bf 319} 188







\bibitem{QGkin} Hu B L 2002 {\it  Int. J. Theor. Phys.} {\bf 41}
pp~2111-38







\end{thebibliography}
\end{document}